\documentclass{nature}

\usepackage[intlimits,sumlimits]{amsmath}
\usepackage{amsfonts,amssymb}
\usepackage{bm}
\usepackage{graphicx}
\usepackage{braket}
\usepackage{hyperref}
\hypersetup{%
pdftitle={Plasmon reflection by topological electronic boundaries in bilayer graphene}, %
pdfauthor={Bor-Yuan Jiang et al.},%
pdfpagemode={UseNone},%
pdfstartview={FitH},%
breaklinks=true,%
citecolor=blue,%
colorlinks=true,%
linkcolor=blue,%
urlcolor=blue
}

\title{Plasmon reflection by topological electronic boundaries in
bilayer graphene}

\author{Bor-Yuan Jiang,$^{1\dagger}$ Guangxin Ni,$^{1\dagger}$
Zachariah Addison,$^2$ Jing K. Shi,$^3$ Xiaomeng Liu,$^3$
Frank Zhao,$^3$ Philip Kim,$^3$
E. J. Mele,$^2$ D. N. Basov,$^{1,4}$
\& Michael M. Fogler$^{1\star}$
}

\begin{document}

\maketitle

\begin{affiliations}
\item Department of Physics, University of California San Diego, La Jolla, California 92093, USA

\item Department of Physics \& Astronomy, University of Pennsylvania, Philadelphia, Pennsylvania 19104, USA

\item Department of Physics, Harvard University, Cambridge, Massachusetts 02138, USA

\item Department of Physics, Columbia University, New York, New York 10027, USA

\item[$^\dagger$] These authors contributed equally to this work.

\item[$^\star$] e-mail: mfogler@ucsd.edu
\end{affiliations}

\begin{abstract}

Domain walls separating regions of $\mathbf{AB}$ and $\mathbf{BA}$ interlayer stacking in bilayer graphene
have attracted attention as
novel examples of structural solitons,\cite{Alden2013sst, Lin2013asb, Butz2013dbg}
topological electronic boundaries,\cite{Zhang2013vcn, Vaezi2013tes, Koshino2013ett, San-Jose2014sbt, Ju2015tvt, Jiang2016sdp, Yin2016dit}
and nanoscale plasmonic scatterers.\cite{Jiang2016sdp}
We show that strong coupling of domain walls to surface plasmons observed in infrared nanoimaging experiments\cite{Jiang2016sdp} is due to topological chiral modes\cite{Zhang2013vcn, Ju2015tvt} confined to the walls.
The optical transitions among these chiral modes and the band continua
enhance the local ac conductivity, which leads to plasmon reflection
by the domain walls.
The imaging reveals two kinds
of plasmonic standing-wave interference patterns,
which we attribute to shear and tensile domain walls.
We compute the 
electron structure of both wall varieties
and show that the tensile wall contains additional confined bands which produce a structure-specific contrast of the local conductivity.
The calculated plasmonic interference profiles
are in quantitative agreement with our experiments.

\end{abstract}

Topological band theory has become a valuable tool
for interpreting ground-state properties and low frequency transport
in electronic materials with nontrivial momentum-space geometry.\cite{Bansil2016ctb} 
In this paper we demonstrate that topological states
may also play a significant role
in the response of such materials at finite frequencies.
Our objects of study are domain walls in bilayer graphene
that separate regions of local $\mathrm{AB}$ and $\mathrm{BA}$ stacking order.
Because of their prevalence in exfoliated samples\cite{Brown2012ttt,
Alden2013sst,
Lin2013asb, Butz2013dbg} and their intriguing electronic\cite{Koshino2013ett,
San-Jose2014sbt, Ju2015tvt} and optical properties,\cite{Jiang2016sdp}
these domain walls have been under intense investigation.\cite{Yin2016dit, Zhang2013vcn, Vaezi2013tes, Koshino2013ett}
From the point of view of the crystal structure,
the stacking wall is a line of partial dislocations.
The magnitude of its Burgers vector is equal to the bond length $|\vec{b}| = 
a / \sqrt{3}$ rather than the lattice constant
$a = 0.246\,\mathrm{nm}$.\cite{Butz2013dbg}
The domain wall can have an
arbitrary angle $\alpha$ with respect to $\vec{b}$,
the two limiting varieties being the tensile wall [$\alpha = 0$,
Fig.~1a] and the shear wall
[$\alpha = \pi / 2$, Fig.~1d].
The width $l = 6$--$10\,\mathrm{nm}$ of a domain wall is determined by a competition between the stacking-dependent interlayer interaction and intralayer elastic strain.\cite{Alden2013sst, Lebedeva2016dsc}
The electronic structure of the domain walls
and their topological properties are best elucidated
using a long-wavelength effective theory\cite{McCann2013epb, Koshino2013ett}
valid for states near the Brillouin zone corners.
In this approach the two inequivalent corners are assigned different valley quantum numbers that are practically decoupled since $l \gg a$. 
Far from the wall, the limit of perfect $\mathrm{AB}$ or $\mathrm{BA}$ stacking is approached.
There, neglecting small ``trigonal warping'' effects,
the dispersion of each valley
consists of parabolic conduction and valence bands touching at a point.
An electric field applied normal to the bilayer
can be used to separate the bands by a tunable energy gap, Figs.~1a and 1d.
However, at the domain wall gapless electron states must remain
since the valley-Chern number of the filled valence band
(of a given valley)
differs by two in the $\mathrm{AB}$ and $\mathrm{BA}$ stacked regions.
Accordingly, the wall must host a minimum of two co-propagating one-dimensional (1D) conducting channels per spin per valley.\cite{Zhang2013vcn, Li2016gct, Martin2008tcb}
In particular, for the tensile wall which runs along a zigzag direction (henceforth, the  -direction), the two valleys project to widely separated conserved momenta parallel to the wall. These 1D channels are protected if one neglects intervalley scattering and so they can support unidirectional valley currents along the wall.

While the above arguments have been used to interpret
dc transport experiments,\cite{Ju2015tvt}
topological considerations alone do not delineate the response at infrared (IR) frequencies $\omega$
where optical transitions between many different electron states
may contribute.
Recent scattering-type scanning near-field optical microscopy (s-SNOM)
experiments\cite{Jiang2016sdp} have demonstrated that in this frequency range the $\mathrm{AB}$-$\mathrm{BA}$ walls act as {\textit {reflectors}} for surface plasmons. Here we show that these reflections probe a structure-sensitive local conductivity of the wall. Specifically, our experiments show that the standing waves formed by the superposition of plasmons launched
by the microscope with plasmons scattered by the domain wall generically depend on the type of wall (i.e., tensile or shear), chemical potential, and the potential bias between the layers.
Our modeling demonstrates that while the charge density is nearly unchanged across a domain wall, the local ac conductivity tensor $\sigma_{i j}(\mathbf{r})$ and thus electrodynamic impedance can be changed significantly because of the presence of both topological and conventional 1D conducting channels.
(Unlike the topological bound states,
the conventional ones do not cross the band gap.)
Incorporating the position-dependent $\sigma_{i j}(\mathbf{r})$ into a long-wavelength theory for the plasmon dispersion, we develop a quantitative description of the observed standing waveform and provide a proof of principle that one can use s-SNOM as a spectroscopic probe of local electronic features at the $10$-$\mathrm{nm}$ scale. This approach should be generally applicable to s-SNOM imaging in a wide family of van der Waals heterostructures.

\begin{figure}
\AtEndDocument{\newpage\includegraphics[width=6 in]{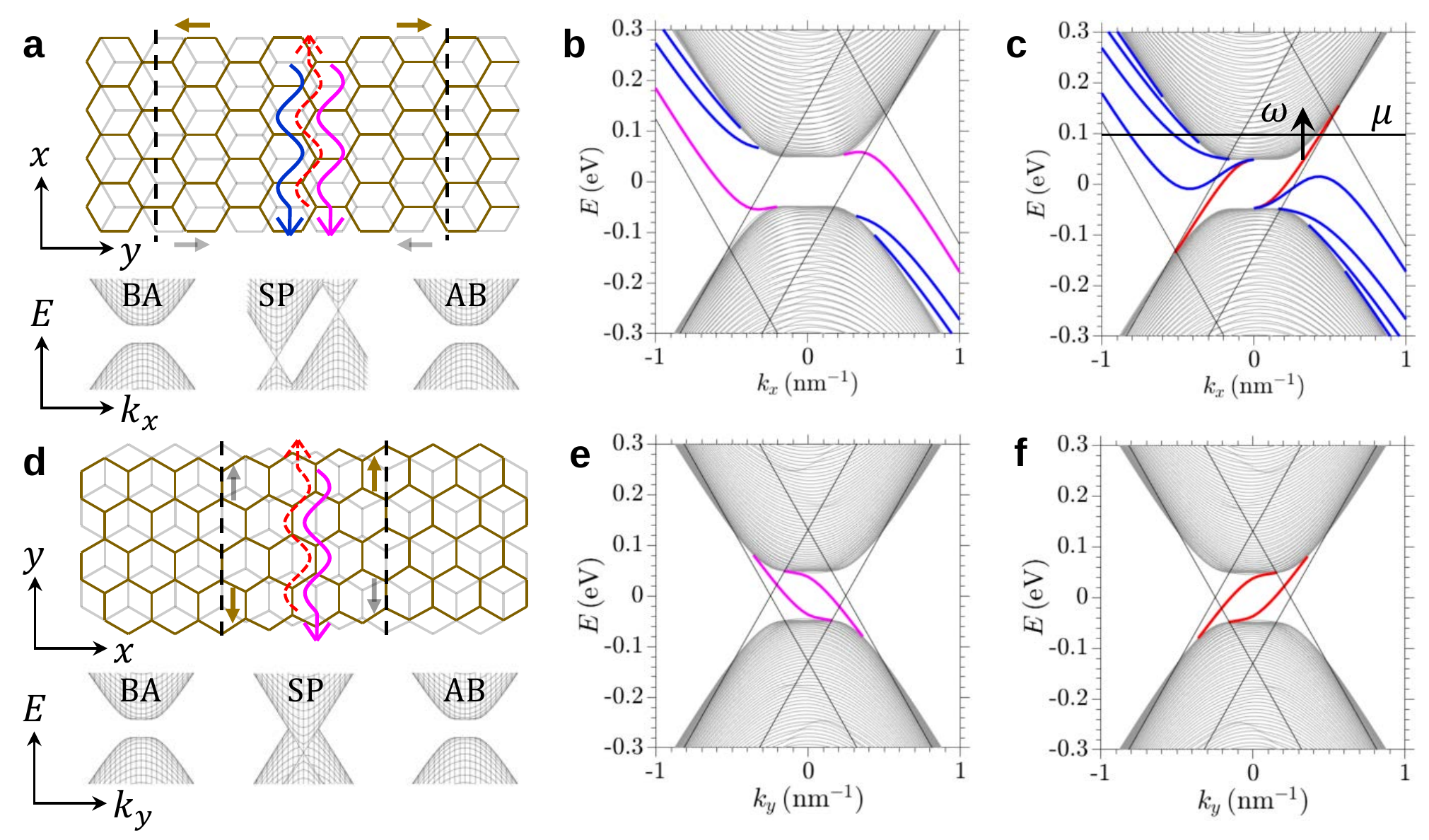}\newline}
\caption{\textbf{Electronic structure of the $\mathbf{AB}$-$\mathbf{BA}$ walls.}
\textbf{a}, Schematic representation of a tensile domain wall and the adiabatic band structures for $\mathrm{BA}$, $\mathrm{SP}$ and $\mathrm{AB}$ stacking.
Brown (gray) arrows indicate the direction of strain for the top (bottom) layer.
Red and magenta (blue) wavy arrows represent the 1D electron states of topological (conventional) origin bound to the domain wall. 
\textbf{b} (\textbf{c}), Band structure of the wall under a positive (negative) interlayer bias $V_i=V_\mathrm{top}-V_\mathrm{bot}$ for the $K$ valley.
The bound states exist outside the continua but roughly within the boundaries of the $SP$ bands indicated by the thin black lines.
The propagation direction of the topological states are fixed by the sign of $V_i$, while for the conventional ones it's fixed by the structure of the $SP$ band and is mostly unchanged by $V_i$.  
\textbf{d}-\textbf{f}, Similar plots for the shear wall.
Due to its smaller width the shear wall hosts no conventional bound states, leading to a distinctly different optical response from the tensile wall.
}
\label{fig1}
\end{figure}

The presence of the extra 1D channels at the domain wall can be understood as arising from a smooth variation of the stacking order across the wall.
If we envision this variation is very gradual, then each point is characterized by a local band structure of bilayer graphene with one layer shifted uniformly relative to the other layer by a certain
$\vec{\delta}=(0, \delta)$.\cite{Koshino2013ett}
In this adiabatic limit the local band structure at the middle of the wall is approximated by $\mathrm{SP}$ (saddle point) stacking, while far away the band structure reverts to that of $\mathrm{AB}$ or $\mathrm{BA}$, as shown
in Figs.~\ref{fig1}1a and 1d for the tensile and shear walls, respectively.
These Figures can be understood as projections from the 2D momentum space to
the 1D momentum axis parallel to the wall:
$k_\parallel = k_x$ for the tensile wall and $k_\parallel = k_y$ for the shear wall.
The 2D $\mathrm{SP}$ dispersion
has two Dirac points shifted in both momentum and energy by the
amount determined by the interlayer hopping amplitude $\gamma_1 = 0.41\,\mathrm{eV}$, the bias voltage $V_i$, and the valley index.
For a tensile wall, these Dirac points 
project to different $k_x$ and remain distinct.
For a shear wall, they occur at the same $k_y$ and overlap with other
states.
In both cases the dispersion at the middle of the wall is gapless.
This is a crucial property
which implies there is a range of energies $E$ in the
$\mathrm{SP}$ stacking that fall into the gap of the $\mathrm{AB}$ stacking.
The states at such energies must be confined to the wall.
These states can be thought of as electronic ``waveguide modes.''
For an infinitely wide waveguide there are infinitely many such states, and the band structure of the entire system is essentially an overlap of the $\mathrm{SP}$ and the $\mathrm{AB}$ bands.
For the realistic situation of finite-width domain walls,
the number of 1D bound states is finite
because the quantum confinement produces a finite number of dispersing 1D branches.
Figures~1b and 1c illustrate this for a positive and negative interlayer bias $V_i$, respectively.
Among all the confined branches, only one pair (per spin per valley) crosses the gap.
These are the topological chiral modes inferred from the valley-Chern number
mismatch mentioned earlier.\cite{Zhang2013vcn, Martin2008tcb}
The propagation direction of these states is determined by the sign of $V_i$ and the valley index.
The remaining confined branches are clearly inherited from the $\mathrm{SP}$ band structure,
in agreement with our qualitative picture.
These waveguide branches have mostly fixed direction regardless of the sign and magnitude of the bias.
Actually, their existence is also a topological effect
to some extent since it is facilitated by the
gaplessness of the $\mathrm{SP}$ dispersion,
which is ensured by the presence of the Dirac points.
The latter is a topological property protected by the spatial inversion and the time-reversal
symmetries.\cite{Koshino2013ett}

The electronic
structure of the shear wall, Fig.~1e,f, can be understood in a similar manner.
The shear wall is narrower than the tensile wall,
so that the quantum confinement effect is stronger.
As a result, the dispersions of the waveguide modes are pushed
extremely close to the
boundaries of the conduction and valence bands.
They are hardly visible in Fig.~1e,f.
Thus, for practical purposes, the waveguide modes are absent and
only the gap-crossing doublet of chiral states survives.

These differences in the band structures for the two types of walls
result in different local optical responses, which we
observed by imaging plasmonic reflections using s-SNOM.
The principles of this experimental technique have been
presented in prior works\cite{Fei2012gtg, Fei2013epp, McLeod2014M, Jiang2016G} and review
articles.\cite{Keilmann2004nfm, Atkin2012n, Basov2014}
In brief,
the s-SNOM utilizes a sharp metallized tip of an atomic force microscope
as a nano-antenna that couples incident IR light to the surface plasmons in the bilayer graphene sheet (Fig.~2e, inset).
These plasmons propagate radially away from the tip and are subsequently reflected by
inhomogeneities, in this case, the $\mathrm{AB}$-$\mathrm{BA}$ wall.
The intensity of the total electric field underneath the tip has
a correction determined
by the interference of the launched and reflected plasmon waves.
The amplitude of the this interference term oscillates as a function
of the distance between the tip and the reflector, with the
period equal to one-half of the plasmon wavelength.
The detection of these interference fringes is made by
measuring the light backscattered by the tip into the far field
and isolating the genuine near-field signal $s$.
For further details of the experimental procedures, see
Methods.

\begin{figure}
\AtEndDocument{\newpage\includegraphics[width=6 in]{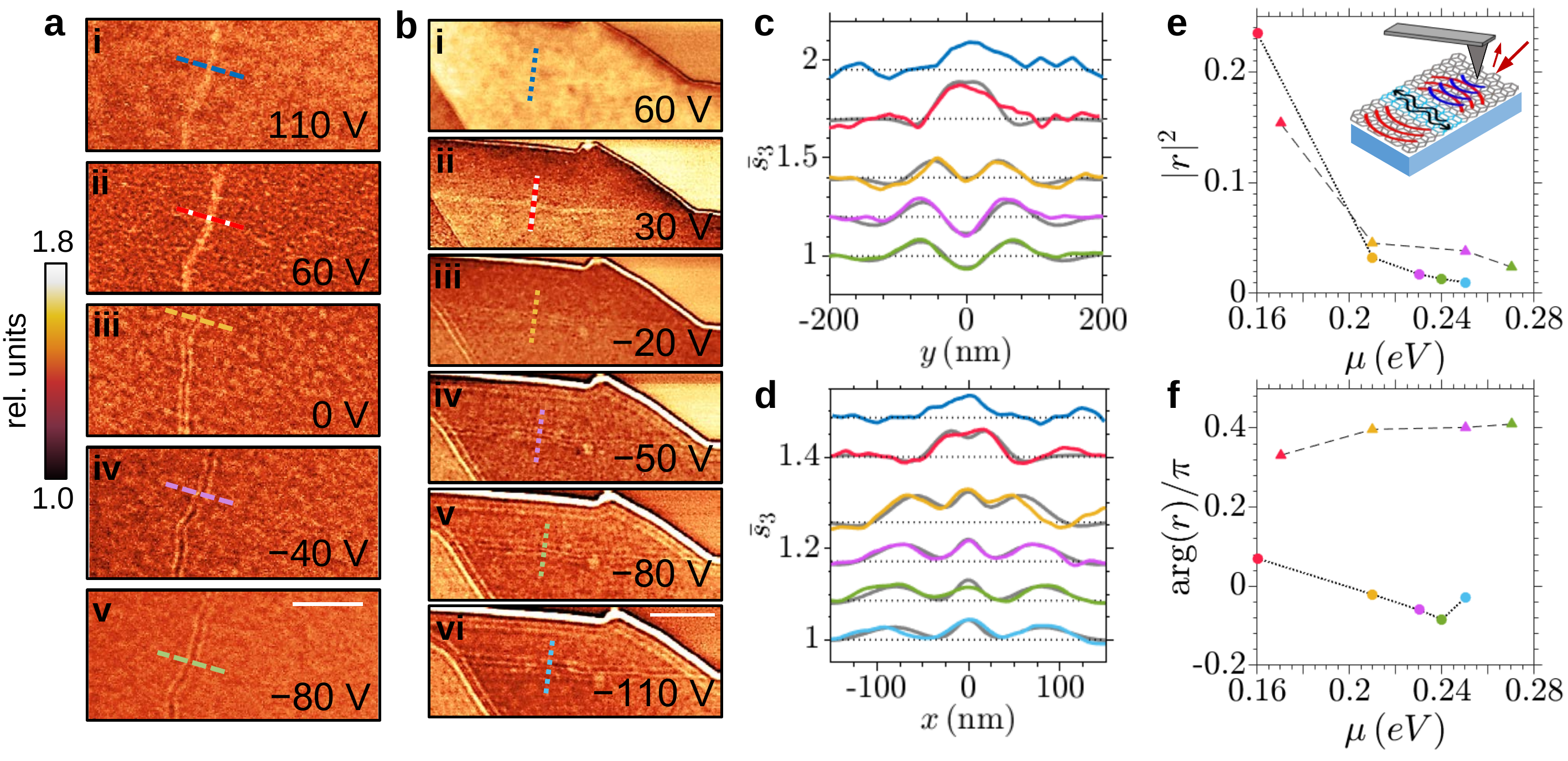}\newline}
\caption{\textbf{s-SNOM images of the $\mathbf{AB}$-$\mathbf{BA}$ walls.}
\textbf{a} (\textbf{b}), Plasmonic interference patterns around a tensile (shear) wall at various gate voltages $V_g$.
Scale bar $1\,\mathrm{\mu m}$.
\textbf{c} (\textbf{d}), Line profiles across the tensile (shear) wall taken at locations indicated by the
colored dashed (dotted) lines in \textbf{a} (\textbf{b}).
Here the s-SNOM amplitude $\bar{s}_3$ is normalized to $1$ at $y=-200\,\mathrm{nm}$ ($x=-150\,\mathrm{nm}$) and offset for clarity,
while fits to the experimental profiles are shown in gray.
As the gate voltage decreases, the amplitude profile across the tensile wall exhibits a smooth transition from one to two peaks, while for the sheer wall the transition is from one to three peaks.
\textbf{e} (\textbf{f}), Magnitude (phase) of the plasmon reflection coefficient $r$ of the domain walls calculated from the fits in \textbf{c} and \textbf{d}.
The triangles represent the tensile wall while the circles represent the shear wall.
For the tensile wall $r$ has a larger magnitude and is mostly imaginary, while for the shear wall $r$ is smaller and mostly real. 
The difference in $r$ between the walls is due to the presence of conventional bound states at the tensile wall in addition to the topological ones.
(Inset) Schematic of the s-SNOM experiment.
The external light source (red arrow) is converted  by the AFM tip into plasmons (blue),
which are partially reflected by the domain wall (cyan) due to the presence of bound states (black wavy arrows). 
The plasmonic interference pattern (blue and red) is detected in the form of back-scattered light (small red arrow).
}
\label{fig2}
\end{figure}

The interference patterns are visualized in the s-SNOM images of
the tensile and the shear wall, shown in Figs.~\ref{fig2}2a and 2b.
These images are acquired from two samples of bilayer graphene, each deposited on a $\mathrm{SiO_2}$ substrate above a $\mathrm{Si}$ global back gate.
For the tensile wall, the pattern is a barely visible bright line when the back gate voltage $V_g$ is close to the charge neutrality point, Fig.~\ref{fig2}2{a}(i).
As $V_g$ decreases,
the pattern evolves into twin interference fringes as shown in Fig.~\ref{fig2}2a(ii-v).
The shear wall behaves similarly, except the pattern starts from a single fringe and evolves into three, Fig.~\ref{fig2}2{b}.
The evolution of the interference patterns can be seen more clearly in the s-SNOM line profiles perpendicular to the wall, Fig.~\ref{fig2}2c,d, where the average of profiles over a $1\,\mu\mathrm{m}$-long section of the wall is shown.

The observed plasmonic scattering and interference patterns are related to
the spatial dependence of the optical conductivity $\sigma_\perp$ of the sheet.
This parameter determines the
momentum $q_p$ of the plasmons
according to the formula $q_p = \frac{i\kappa\omega}{2\pi\sigma_\perp}$
where $\kappa$ is the effective
permittivity of the environment.\cite{Basov2014}
Depending on the type of wall and their respective local electronic structure, the perturbation of local $q_p$ by the wall will vary, leading to distinct plasmonic signatures for each type of wall.
Additionally, as the gate voltage $V_g$ is tuned,
the carrier density and chemical potential of the bilayer graphene sheet are tuned correspondingly.
By tuning $V_g$ and studying the corresponding s-SNOM signal, one effectively probes different parts of the electronic spectrum.

For a quantitative analysis, we calculate the local optical conductivity of the domain wall following these steps.
We start with a model\cite{Koshino2013ett} of $4\times 4$ Dirac-type
Hamiltonian $H = H(k_\parallel, k_\perp)$
of bilayer graphene with a uniform arbitrary stacking $\delta$.
We modify it by allowing smooth spatial variations of the stacking parameter $\delta
= \delta(x_\perp)$.
Note that $x_\perp=y$ for the tensile wall and $x_\perp=x$
for the shear wall where $x$ is the zigzag direction of the graphene lattice,
see Fig.~1.
The momentum $k_\parallel$ remains a good quantum number.
The momentum $k_\perp$ perpendicular to the wall is replaced by
the operator $-i\partial/\partial x_\perp$.
The resulting eigenproblem is solved numerically on a 1D grid of $x_\perp$ to obtain electronic band dispersions such as those shown in Fig.~\ref{fig1}1.
Next, these energy dispersions and the wave functions are used to
evaluate the Kubo formula for the nonlocal conductivity $\Sigma(x_\perp,x_\perp')$.
We further define an effective local conductivity $\sigma_\perp(x_\perp)\equiv\int \Sigma(x_\perp,x_\perp')dx_\perp'$, which is appropriate when the total electric field on the sheet varies slowly on the scale of the wall width.\cite{Jiang2016tpr}
The presence of the bound states is manifest
in a strongly enhanced
$\sigma_\perp$ at the tensile domain wall.
The  profile of $\sigma_\perp(x_\perp)$ obtained from this nonlocal
model can be compared with
the results for an adiabatic model in which
the Hamiltonian for the uniform stacking is used instead at every point $x_\perp$.
As shown in Fig.~3a,
the real part of $\sigma_\perp$ is much larger than that obtained from the adiabatic approximation.
The difference occurs precisely because of the lack of
the bound states in the latter.
On the other hand, for shear walls which host no bound states when $V_i=0$, the conductivity is relatively flat at the wall, see Fig.~3b.

The large contrast of the local optical conductivity for the two types of walls leads directly to the distinct plasmonic profiles they produce.
To simulate the measured s-SNOM profile, we converted $\sigma_\perp$ into the plasmon momentum $q_p$ and used it as input to an electromagnetic solver we developed in previous work.\cite{Fei2012gtg, Fei2013epp, Jiang2016tpr,Ni2015pgm,Ni2016uos,Goldflam2015tps}
The chemical potential $\mu$, interlayer bias $V_i$ and the phenomenological damping rate $\eta$ are tuned until the output matches the experimental profile.
An example of a fit for the tensile wall is shown in Fig.~4a-d and for the shear wall in Fig.~4e-h, where the magnitude and phase of the simulated s-SNOM signal are shown.
Here the plasmon momentum $q_p$ is parametrized by the plasmon wavelength $\lambda_p$ and plasmonic damping rate $\gamma$, $q_p=\frac{2\pi}{\lambda_p}(1+i\gamma)$.
The interaction of the walls with plasmons can be characterized by a single parameter, the plasmon reflection coefficient $r$,
as the walls have a width an order of magnitude smaller than the plasmon wavelength and can be regarded as 1D objects.
According to the first-order perturbation theory,\cite{Fei2013epp}
this reflection coefficient is proportional to the amount of excess conductivity at the wall with respect to the background value $\sigma_\infty$ far from the wall: $r=i q_\infty \int dx_\perp(\sigma_\perp(x_\perp)-\sigma_\infty)/\sigma_\infty$.
As shown in Figs~2e and 2f, at higher $\mu$ the tensile wall has a larger and mostly imaginary $r$, while the shear wall has smaller and mostly real $r$.
Therefore, the tensile wall reflects plasmons more strongly, while the phase of $r$ dictates that the interference pattern has a minimum at the tensile wall and a maximum at the shear wall as observed in the experiments.
Note that we are treating the interlayer bias $V_i$ as a fitting parameter
because in the experiment the value of $V_i$ is determined primarily by (uncontrolled) dopants from the substrate.\cite{Zhang2008des}
However, the calculated s-SNOM profiles are strongly $V_i$-dependent because
the sign and value of $V_i$ can alter the bound state dispersions (Fig.~1b and 1c) and the conductivity $\sigma_\perp$ (Fig.~3c and 3d) dramatically.
This suggests that determination of $V_i$ from the fits to the
s-SNOM profiles is a reliable procedure.
Indeed, the remarkable fidelity of our fits which accurately reproduce the measured features in the amplitude profiles across a variety of gate voltages indicates that s-SNOM can be a robust method for probing the local band structure of ultra small electron materials and structures.

One may wonder whether domain walls can support confined plasmons in addition to confined single-particle states. A major obstacle to the existence of such 1D plasmons is Landau damping due to the surrounding bulk continua. This type of damping can be diminished if two conditions are satisfied:
i) the chemical potential $\mu$ resides inside the band gap
so that the confined 1D modes are the only low-energy
degrees of freedom and ii) the
frequency $\omega$ is small enough so that the optical transitions to
bulk bands do not occur.
Devices where these requirements can be met
would likely need a thin top gate in addition to the back gate
to control $\mu$
and also substrates/gate dielectrics other than
SiO$_2$ to reduce disorder and unintentional doping.
The 1D plasmon mode may then be amenable for study by the s-SNOM nanoimaging,
similar to plasmons in carbon nanotubes.\cite{Shi2015oll}
The dispersion of the domain-wall plasmon is determined by the
divergences of the loss function $-\mathrm{Im}\,
\varepsilon_{\mathrm{1D}}^{-1}\left(k_\parallel, \omega\right)$.
Here $\varepsilon_{\mathrm{1D}}$ is the effective dielectric function\cite{Kane1997cim}
\begin{equation}
\varepsilon_{\mathrm{1D}}\left(k_\parallel, \omega\right) = 
\kappa - \frac{8 e^2}{h}\, \ln \left(\frac{A}{k_\parallel l}\right)
\sum_{j = 1}^{N}
\frac{k_\parallel^2 |v_{j}|}{\omega^2 - k_\parallel^2 v_{j}^2}\,,
\label{eqn:1D_plasmon}
\end{equation}
$\kappa$ is the dielectric constant of the environment,
$N$ is the total number (per spin per valley) of
the 1D electron states,
$v_{j}$ is the velocity of $j$th state, and
$A \sim 1$ is a numerical coefficient.
The dispersion
curves calculated using Eq.~\eqref{eqn:1D_plasmon}
(cf.~Supplementary Material for details)
are approximately linear in the experimentally accessible
range of momenta $k_\parallel$, see Fig.~5a.
The slope of each curve, i.e., the plasmon
group velocity scales as $\sqrt{N}$, similar to
the case of carbon nanotubes.\cite{Shi2015oll}
For shear walls only
$N = 2$ is possible (Figs.~1e and 1f), and so
the plasmon wavelength $\lambda_p = 2\pi / k_\parallel$
as a function of $\mu$ at given fixed $\omega$ and $V_i$ is
approximately constant, see the red and green curves in Fig.~5b.
For tensile walls, we can have $N = 2$, $4$, or $6$,
depending on the chemical potential and interlayer bias
(Fig.~5b, inset).
Sharp changes in
$\lambda_p$ should therefore occur at some $\mu$ where $N$ changes in steps of two,
see the blue curve in Fig.~5b.
These properties of plasmons
propagating along  $\mathrm{AB}$-$\mathrm{BA}$ boundaries
may be interesting for
exploring fundamental physics of interacting 1D systems
(so-called Luttinger liquids\cite{Kane1997cim})
or for implementing ultrasmall plasmonic circuits.

In summary, the local electrodynamic impedance 
of bilayer graphene 
is strongly sensitive to the atomic-scale stacking order. This provides a boundary condition for the 
propagation of surface plasmons.
Topological arguments give an important initial insight into the
physics of plasmon reflection by
the domain walls. However,
further microscopic analysis of the electron structure
proves to be necessary to account quantitatively  
for the plasmon interference fringes
observed in near-field nanoimaging.
We anticipate many useful applications of these
theoretical concepts and experimental approaches to
other types of electronic boundaries found
in a wide family of van der Waals heterostructures.

\begin{figure}
\AtEndDocument{\newpage\includegraphics[width=3.5 in]{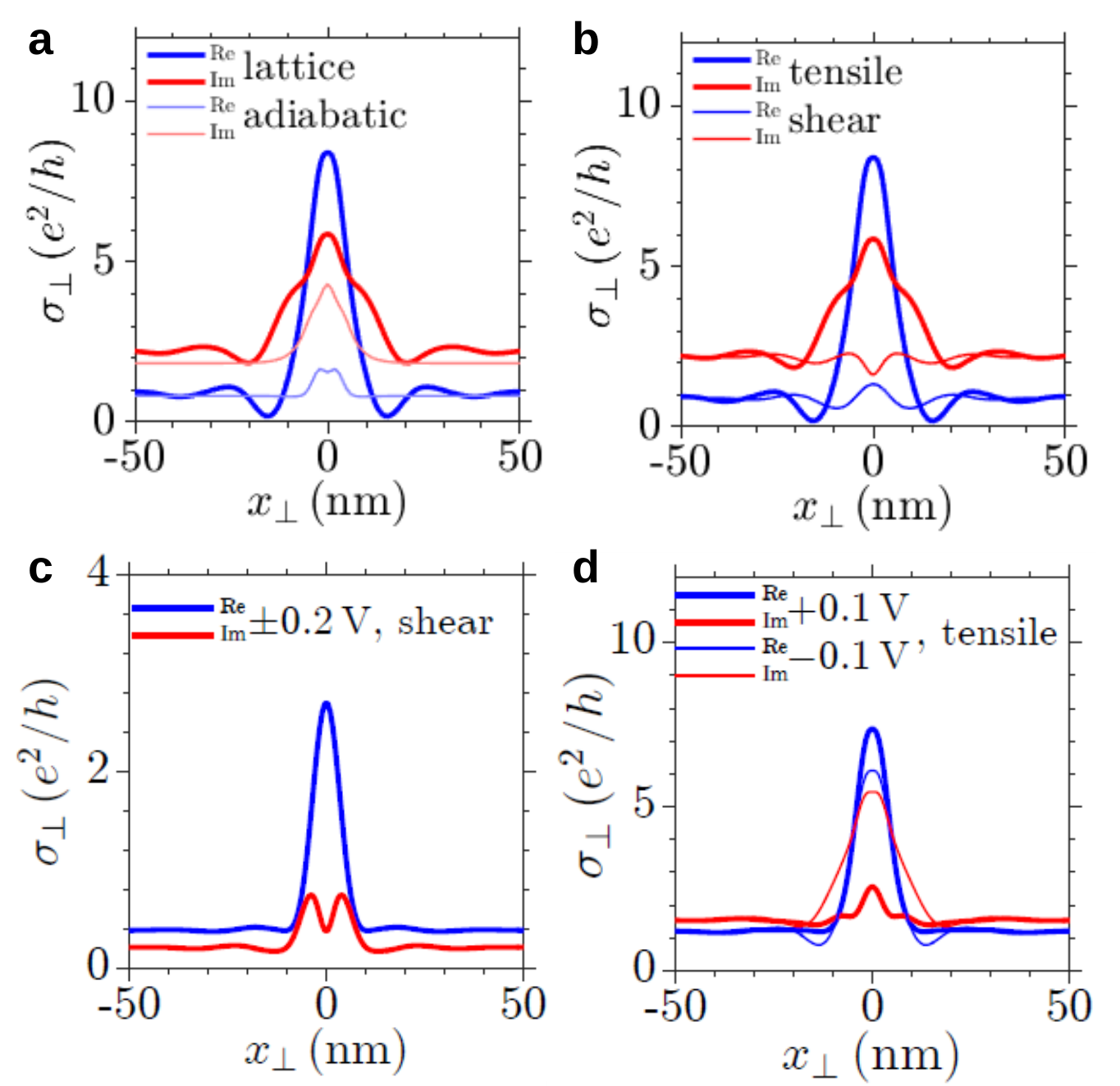}\newline}
\caption{\textbf{Local optical conductivity of the domain walls at $\bm{\mu=0.1}\,\mathbf{eV}$, $\bm{\eta=0.1}$ and $\bm{\omega=890}\,\mathbf{cm^{-1}}$.}
\textbf{a}, Local optical conductivity $\sigma_{\perp}$ for the tensile wall at $V_i=0$, calculated using either the lattice approach or the adiabatic approach. 
The prominent peak in the real part of $\sigma_{\perp}$ for the lattice approach comes from the inclusion of (conventional) bound states, which cannot be account for in the adiabatic approach.
\textbf{b}, Comparison of $\sigma_{\perp}$ between the tensile wall and the shear wall under the same parameters.
At zero bias the shear wall hosts no bound states and the conductivity is relatively flat.
\textbf{c},
Conductivity of the shear wall at interlayer bias $V_i=\pm 0.2\,\mathrm{V}$.
The bias opens the band gap and introduces the topological bound states to the shear wall,
leading to a conductivity that is no longer flat but has a peak in $\mathrm{Re}\,\sigma_{\perp}$ instead.
Note that the sign of the bias does not affect $\sigma_{\perp}$, as reversing the bias is equivalent to interchanging the valleys.
\textbf{d}, 
Conductivity of the tensile wall at $V_i=\pm0.1\,\mathrm{eV}$.
For the tensile wall, the sign of the bias alters the dispersion of the conventional bound states, leading to very different conductivities.
}
\label{fig3}
\end{figure}

\begin{figure}
\AtEndDocument{\newpage\includegraphics[width=4 in]{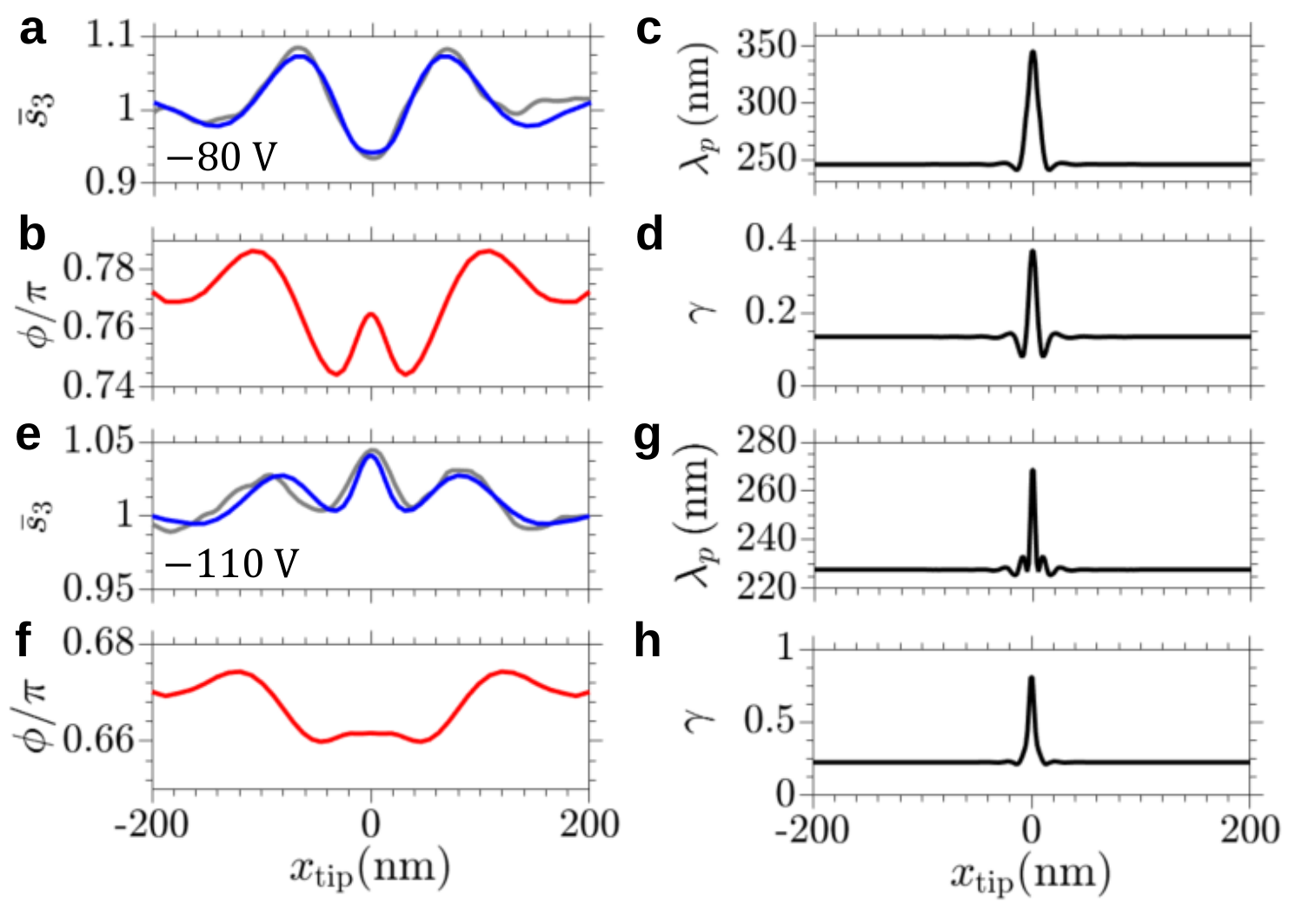}\newline}
\caption{\textbf{Fitting s-SNOM profiles.}
\textbf{a} (\textbf{b}), Simulated s-SNOM amplitude (phase) profiles for the tensile wall at $V_g=-80\,\mathrm{V}$.
Experimental data is shown in gray in \textbf{a}.
\textbf{c} (\textbf{d}), The plasmon wavelength (damping) profile used for the fit.
Peaks in the profiles arise from optical transitions involving the bound states, indicating their importance in determining the optical response around the domain walls.
\textbf{e}-\textbf{h}, Similar plots for the shear wall at $V_g=-110\,\mathrm{V}$.
}
\label{fig4}
\end{figure}

\begin{figure}
\AtEndDocument{\newpage\includegraphics[width=5 in]{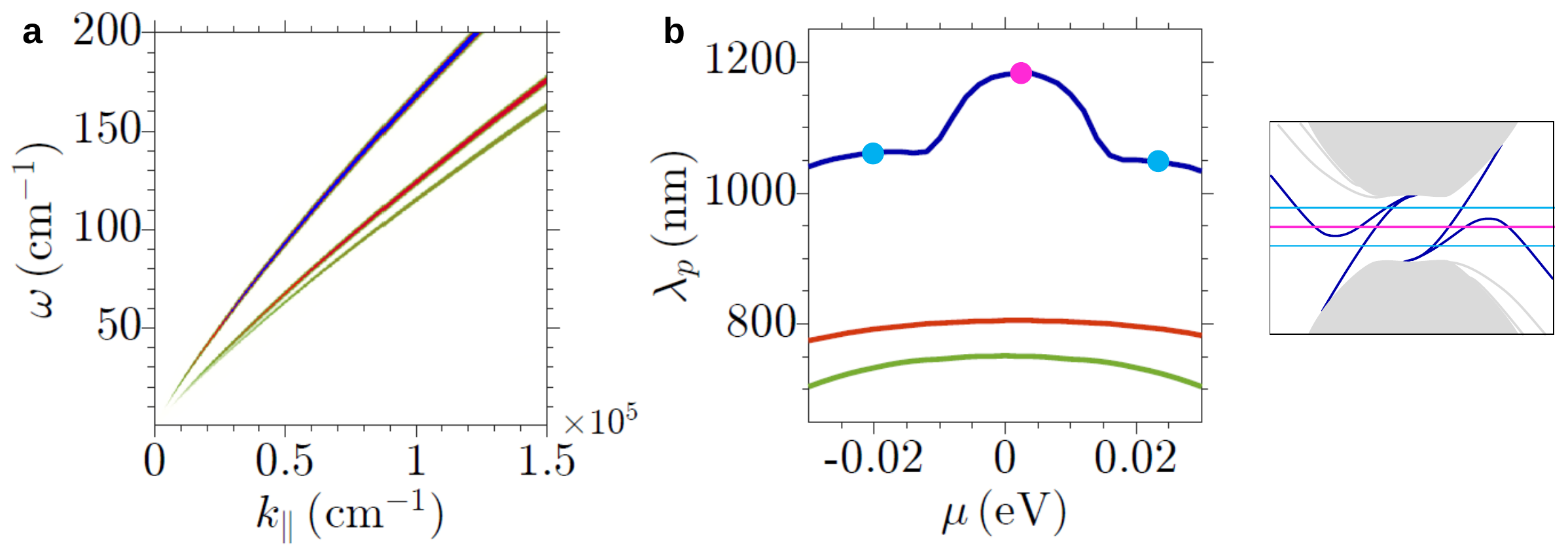}\newline}
	\caption{\textbf{1D plasmons at the wall.}
	\textbf{a}, Dispersion of 1D plasmons at $\mu=0$ for \textbf{i}. the shear wall at $V_i=0.1\,\mathrm{V}$ (green, $N=2$), \textbf{ii}. the tensile wall at $V_i=0.1\,\mathrm{V}$ (orange, $N=2$) and \textbf{iii}. the tensile wall at $V_i=-0.1\,\mathrm{V}$ (blue, $N=6$).
	The difference in the plasmon velocity $v_p=\partial\omega/\partial k_\parallel$ is due to the different number of plasmonic channels $N$, $v_p\propto\sqrt{N}$.
	\textbf{b}, 
	(Left panel) Plasmon wavelength for the three cases in \textbf{a} at $\omega=100\,\mathrm{cm^{-1}}$.
	For case \textbf{iii} the plasmon wavelength changes sharply at particular $\mu$'s due to a change in $N$.
	(Right panel) Corresponding band structure for case \textbf{iii}, where $N$
	varies between $4$ (cyan) and $6$ (magenta) at different chemical potentials $\mu$.
}
\label{fig5}
\end{figure}

\begin{methods}
Graphene flakes were exfoliated onto a $285\,\mathrm{nm}$-thick $\mathrm{SiO_2}$ layer on top of a highly doped $\mathrm{Si}$ substrate. 
Regions of bilayer graphene were identified by their contrast under optical microscopy. 
Metal contacts were defined on graphene using shadow masks.  
The infrared nanoimaging experiments were performed
at ambient conditions using an s-SNOM based on an atomic force microscope operating in the tapping mode.
Infrared light ($\lambda = 11.2\,\mathrm{\mu m}$) was focused onto the tip of the microscope. 
A pseudo-heterodyne interferometric detection was used to extract the scattering amplitude $s$ and phase $\phi$ of the near-field signal. 
To remove the background, the signal
was demodulated at the third harmonic of the tapping frequency
$270\,\mathrm{kHz}$. 
\end{methods}



\bibliographystyle{naturemag}
\bibliography{abba_plasmon}


\begin{addendum}
\item This work is supported by the DOE under
 Grant DE-SC0012592,
 by the ONR under Grant N00014-15-1-2671 and N00014-15-1-2761,
 by the NSF under Grant ECCS-1640173,
 and by the SRC.
 D.N.B. and P.K. are investigators in Quantum Materials funded by
 the Gordon and Betty Moore Foundation's EPiQS Initiative
 through Grant No. GBMF4533  and GBMF4543.
 Work by Z.A. and E.J.M. was supported by the DOE under grant DE FG02 84ER45118.

\item[Author contributions] 
D.N.B. and E.J.M. conceived the project.
B.-Y.J., Z.A., E.J.M. and M.M.F. developed the theoretical model.
G.N. performed the experiments and analysed the data.
J.K.S., X.L., F.Z. and P.K. prepared the samples.
All authors contributed to the manuscript.

\item[Additional information] Correspondence and requests for materials
should be addressed to M.M.F.

\item[Competing financial interests] The authors declare that they have no
competing financial interests.

\end{addendum}
\end{document}


\maketitle

\begin{affiliations}
	\item Department of Physics, University of California San Diego, La Jolla, California 92093, USA
	
	\item Department of Physics \& Astronomy, University of Pennsylvania, Philadelphia, Pennsylvania 19104, USA
	
	\item Department of Physics, Harvard University, Cambridge, Massachusetts 02138, USA
	
	\item Department of Physics, Columbia University, New York, New York 10027, USA
	
	\item[$^\dagger$] These authors contributed equally to this work.
	
	\item[$^\star$] e-mail: mfogler@ucsd.edu
\end{affiliations}

\section{Model for bilayer graphene}

Our low energy four-band Hamiltonian for homogeneous BLG is adopted from Ref.~\onlinecite{Koshino2013ett},
\begin{equation}
\bar{H}=
\begin{bmatrix}
H_{0}^+ & U^\dagger\\
U & H_{0}^-
\end{bmatrix}
\,,
\label{eqn:H}
\end{equation}
with basis $(F_{At}^{K^\xi},F_{Bt}^{K^\xi},F_{Ab}^{K^\xi},F_{Bb}^{K^\xi})$, where $F$ denotes envelope function, $\xi=\pm 1$ for the $K$ and $K'$ valley, and $t$ stands for top layer and $b$ for bottom layer.
The two-band Dirac Hamiltonian for a single layer is
\begin{equation}
H_0^\pm=
\begin{bmatrix}
\pm V/2 & \xi k_x + ik_y\\
\xi k_x - ik_y & \pm V/2
\end{bmatrix}
\,,
\end{equation}
where $V=eV_i$ is the interlayer potential. 
The interlayer interaction is
\begin{equation}
U=\frac{\gamma_1}{3}\left(1+2
\begin{bmatrix}
\cos \frac{2\pi}{3}\delta & \cos \frac{2\pi}{3}\left(\delta+1\right)\\
\cos \frac{2\pi}{3}\left(\delta-1\right) & \cos \frac{2\pi}{3}\delta
\end{bmatrix}
\right)\,,
\end{equation}
where the interlayer coupling\cite{Zhang2008des} $\gamma_1=0.4\,\mathrm{eV}$ and the stacking order $\delta\in [1,2]$ with $\delta=1,1.5,2$ corresponding to $AB$, $SP$ and $BA$ stacking.
To describe the domain walls, 
the homogeneous Hamiltonian $\bar{H}$ has to be modified to account for the local change in stacking.
This is done by replacing the momentum perpendicular to the wall $k_\perp$ by the operator $-i\partial/\partial x_\perp$ and making the stacking parameter spatially dependent, $\delta(x_\perp)$,
resulting in the real space Hamiltonian $H(x_\perp)$.
For the tensile wall $x_\perp=y$ while for the shear wall $x_\perp=x$.
The distribution $\delta(x_\perp)$ is found in Ref.~\onlinecite{Alden2013sst} to be
\begin{equation}
\delta(x_\perp)=\frac2\pi\arctan\left(e^{\pi{x_\perp}/{l}}\right)+1\,,
\end{equation}
where the width $l=10.1\,\mathrm{nm}$ for the tensile wall and $l=6.2\,\mathrm{nm}$ for the shear wall.

\section{Optical conductivity of the domain wall}

There are two ways to approximate $\sigma(x_\perp)$.
The first is to diagonalize the homogeneous Hamiltonian $\bar{H}$ for a given stacking order $\delta$,
use the Kubo formula to find the homogeneous optical conductivity $\bar{\sigma}(\delta)$,
then map it to $\bar{\sigma}(x_\perp)$ using the stacking distribution $\delta(x_\perp)$ of the domain wall. 
This is what we call the ``adiabatic'' approach and cannot account for the presence of the edge states.
The second is to diagonalize the real space Hamiltonian $H$ in coordinate basis and use the Kubo formula to find the nonlocal conductivity $\Sigma(x_\perp,x_\perp')$, which is then localized by $\sigma(x_\perp)=\int \Sigma(x_\perp,x_\perp')dx_\perp'$.
This ``lattice'' approach is what we use for our calculations. 

Let us start from the calculation of $\bar{\sigma}(\delta)$.
The conductivity consists of two parts, an interband conductivity $\bar{\sigma}^I$ from optical transitions between the four bands, and a Drude-like intraband conductivity $\bar{\sigma}^D$. 
Except for specific stacking orders such as $AB$ and $BA$ stacking, the conductivities are anisotropic.
We consider only the diagonal elements of the conductivity, $\bar{\sigma}_{xx}$ and $\bar{\sigma}_{yy}$, and neglect $\bar{\sigma}_{xy}$ and $\bar{\sigma}_{yx}$ which are small.\cite{Shimazaki2015gdp}
The interband conductivity is calculated using the Kubo formula,
\begin{equation}
\bar{\sigma}^I_{\alpha\alpha}=\frac{g_s g_v i \hbar}{4\pi^2}\int dk_xdk_y\sum_{n\neq m}-\frac{f_m-f_n}{E_m-E_n}\frac{e^2v^2M_\alpha^*M_\alpha}{\hbar\omega(1+i\eta)-(E_m-E_n)}\,.
\label{eqn:kubo}
\end{equation}
Here $\alpha=x$ or $y$. 
The spin and valley degeneracy are $g_s=g_v=2$. 
The summation goes over all pairs of states $\ket{n}$ and $\ket{m}$,
where the energy of the state $\ket{n}$ is $E_n$ and its occupation number $f_n$ is given by the Fermi-Dirac distribution, $f_n=1/(1+e^{(E_n-\mu)/k_BT})$.
The matrix element is defined as $M_\alpha=\braket{m|s_\alpha\otimes\tau_0|n}$ where $s_\alpha$ are the Pauli matrices acting on the sublattice and $\tau_0$ is the identity matrix acting on the layer degree of freedom.
The phenomenological damping rate is $\eta$.

The intraband conductivity $\bar{\sigma}^D$ arises from the $n=m$ part of the summation,
where the fraction $\frac{f_m-f_n}{E_m-E_n}$ is replaced by the derivative $\frac{df_n}{dE_n}$ while $E_m-E_n=0$, so that
\begin{equation}
\bar{\sigma}^D_{\alpha\alpha}=\frac{g_s g_v i \hbar}{4\pi^2}\int dk_xdk_y\sum_{n}-\frac{df_n}{dE_n}\frac{e^2v^2M_\alpha^*M_\alpha}{\hbar\omega(1+i\eta)}\,.
\label{eqn:kubo_D}
\end{equation}
The total conductivity $\bar{\sigma}=\bar{\sigma}^I+\bar{\sigma}^D$
can be readily found given the Hamiltonian $\bar{H}(\delta)$, the chemical potential $\mu$, the temperature $T$, the frequency $\omega$, the interlayer bias $V_i$ and the damping rate $\eta$.

\begin{figure}[t]
	\begin{center}
		\includegraphics[width=2.1 in]{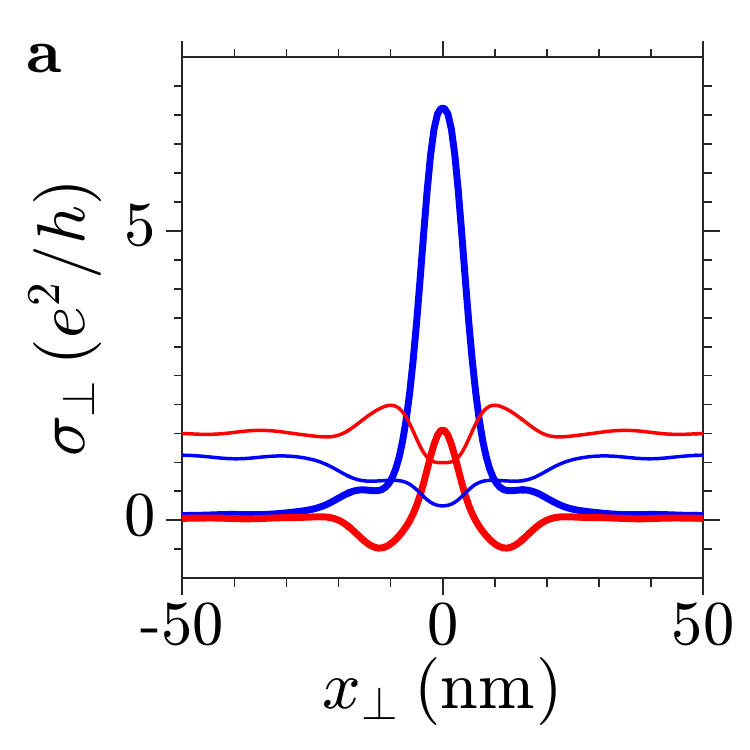}
		\includegraphics[width=2.1 in]{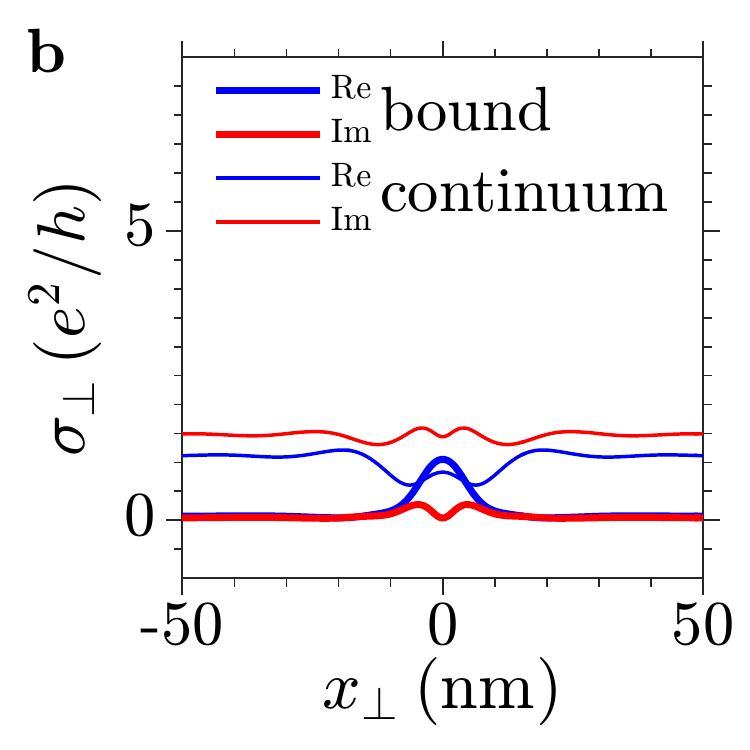}
	\end{center}
	\caption{
		\textbf{a}. Local conductivity $\sigma_\perp$ for the tensile wall,
		where the contribution from optical transitions involving the bound states (thick curves) are separated from the contribution of transitions involving only the continuum (thin curves).
		Parameters: $\mu=0.1\,\mathrm{eV}$, $T=300\,\mathrm{K}$, $\omega=890\,\mathrm{cm^{-1}}$, $V_i=0.1\,\mathrm{V}$ and $\eta=0.1$.	
		\textbf{b}. Similar quantities for the shear wall.
		In both cases the bound states produce a prominent peak in the real part of $\sigma_\perp$.
	}
	\label{fig:sigma_bc}
\end{figure}

The calculation of the nonlocal conductivity $\Sigma$ is very similar.
The system is discretized in the $x_\perp$ direction into a grid of size $N$, so that the Hamiltonian $H$ has $4N$ bands.
The integration over $k_\perp$ is removed, and the matrix element is calculated at every grid point, $M_\alpha(x_\perp)=\braket{m(x_\perp)|s_\alpha\otimes\tau_0|n(x_\perp)}$, leading to the following nonlocal
conductivities
\begin{equation}
{\Sigma}^I_{\alpha\alpha}(x_\perp,x_\perp')=\frac{g_s g_v i \hbar}{4\pi^2}\int dk_\parallel\sum_{n\neq m}-\frac{f_m-f_n}{E_m-E_n}\frac{e^2v^2M_\alpha^*(x_\perp)M_\alpha(x_\perp')}{\hbar\omega(1+i\eta)-(E_m-E_n)}\,,
\label{eqn:kubo_nonlocal}
\end{equation}
\begin{equation}
{\Sigma}^D_{\alpha\alpha}(x_\perp,x_\perp')=\frac{g_s g_v i \hbar}{4\pi^2}\int dk_\parallel\sum_{n}-\frac{df_n}{dE_n}\frac{e^2v^2M_\alpha^*(x_\perp)M_\alpha(x_\perp')}{\hbar\omega(1+i\eta)}\,.
\label{eqn:kubo_D_nonlocal}
\end{equation}
For our calculations the nonlocal $\Sigma$ is then localized by integration over $x_\perp'$ and denoted $\sigma_\alpha\equiv\sigma_{\alpha\alpha}$, where $\alpha=\perp$ or $\parallel$.

\begin{figure}[ht]
	\begin{center}
		\includegraphics[width=2.1 in]{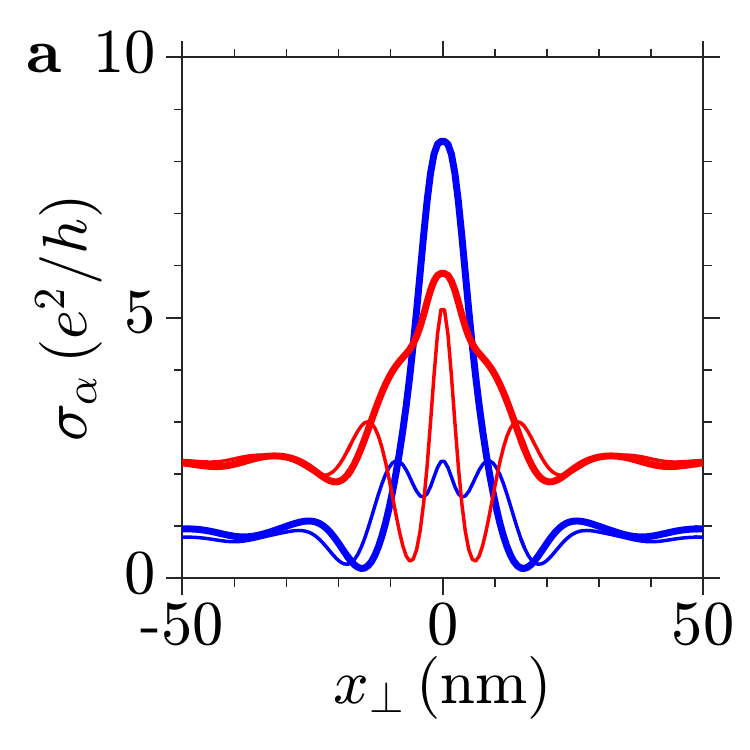}
		\includegraphics[width=2.1 in]{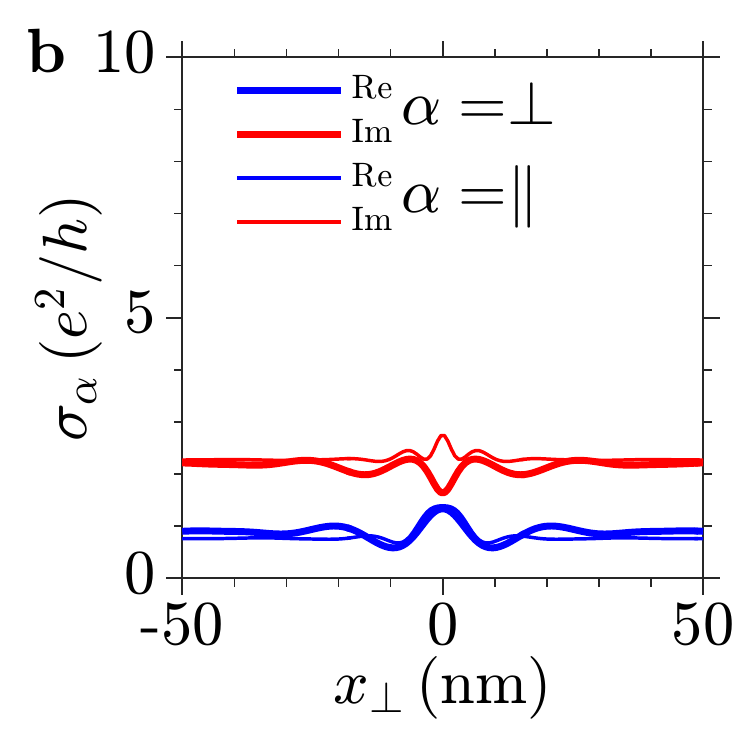}
	\end{center}
	\caption{
		\textbf{a}. The local conductivity $\sigma_\alpha$ is highly anisotropic at the tensile wall.
		Parameteres: $\mu=0.1\,\mathrm{eV}$, $T=300\,\mathrm{K}$, $\omega=890\,\mathrm{cm^{-1}}$, $V_i=0$ and $\eta=0.1$.	
		\textbf{b}. Similar quantities for the shear wall.
	}
	\label{fig:sigma_aniso}
\end{figure}

As the bound state wavefunctions are localized at the domain wall, optical transitions involving these bound states give rise to conductivity peaks at the wall, as shown in Fig.~\ref{fig:sigma_bc}.
The domain wall also introduces anisotropy to the local conductivity,
as shown in Fig.~\ref{fig:sigma_aniso}a for the tensile wall and \ref{fig:sigma_aniso}b for the shear wall.
Away from the wall, conductivities in the $\perp$ and the $\parallel$ direction have the same value as expected, but at the wall they can be drastically different.

\section{Fitting the s-SNOM profiles}

\begin{figure}[t]
	\begin{center}
		\includegraphics[width=2.1 in]{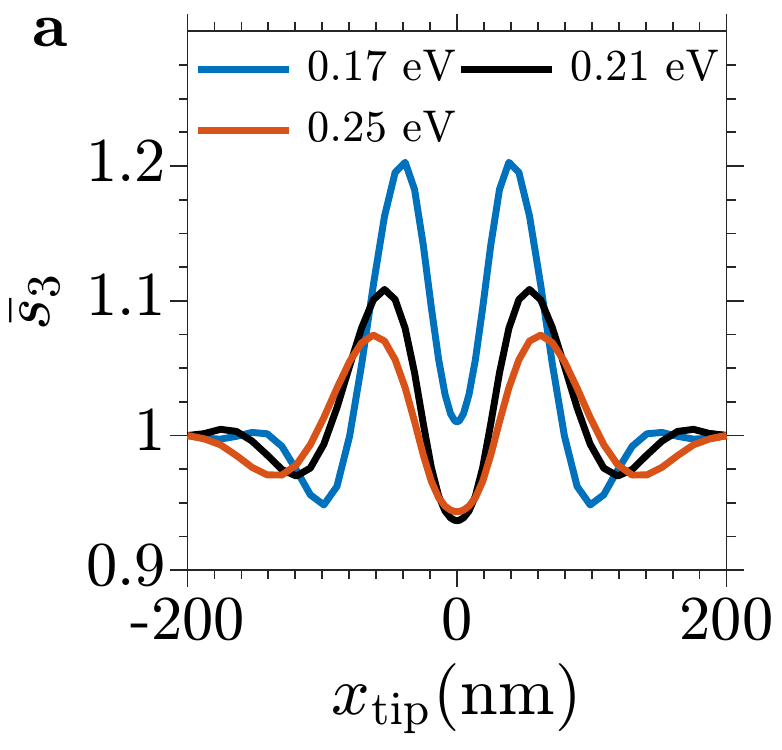}
		\includegraphics[width=2.1 in]{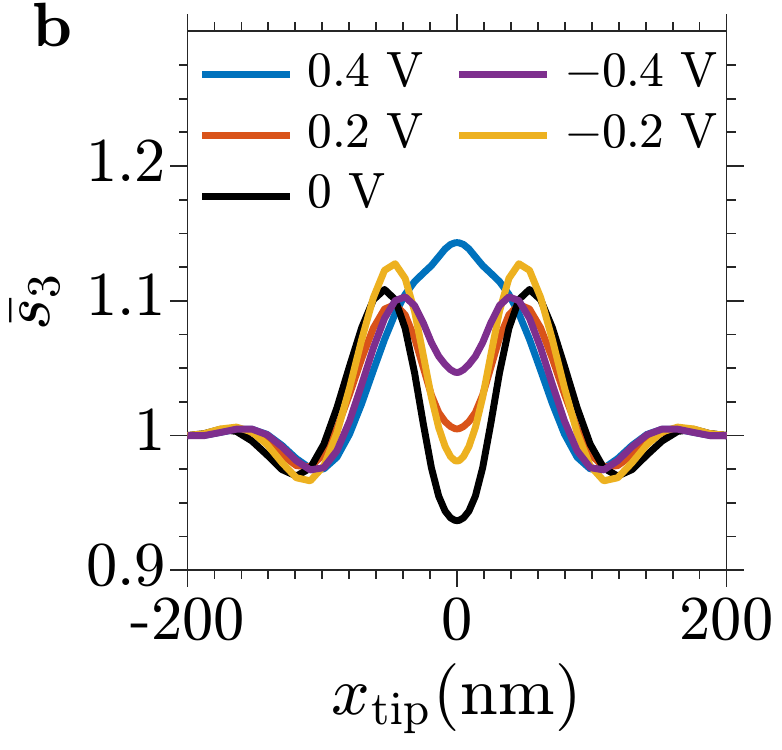}
		\includegraphics[width=2.1 in]{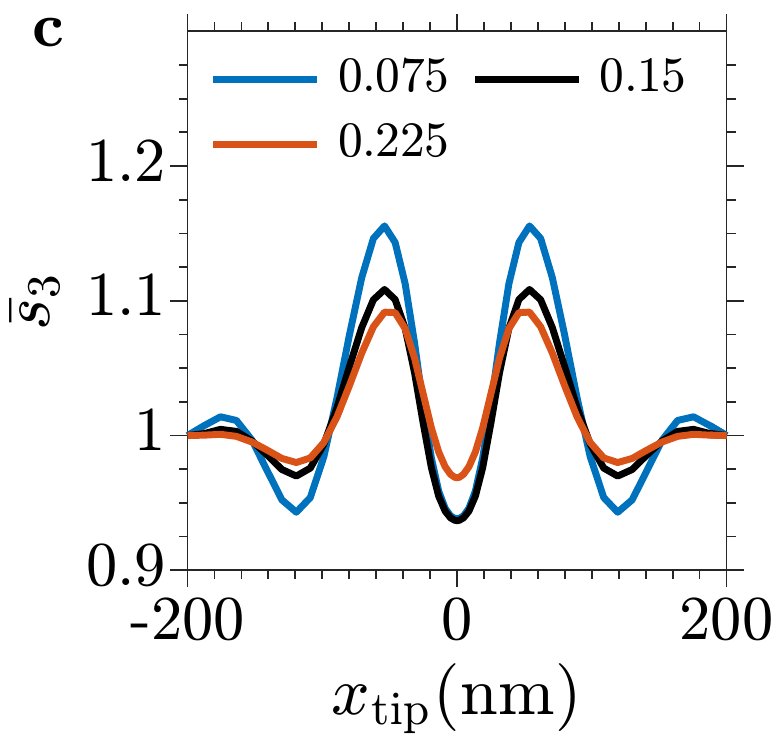}
	\end{center}
	\caption{Comparison of near-field amplitude profiles under different fitting parameters.
	The black curve in every panel is calculated at $\mu=0.21\,\mathrm{eV}$, $V_i=0\,\mathrm{V}$, and $\eta=0.15$ for the tensile wall.
	In each panel one of the three parameters is varied.
	\textbf{a}. Varying the chemical potential changes the plasmon wavelength and the overall amplitude.
	\textbf{b}. Changing the interlayer bias $V_i$ alters signal strength at the wall.
	\textbf{c}. Increasing the damping rate $\eta$ decreases the overall amplitude of the oscillations.
	}
	\label{fig:SNOM_parameters}
\end{figure}

To fit the experimental s-SNOM profiles, we calculate conductivities at $T=300\,\mathrm{K}$ and $\omega=890\,\mathrm{cm^{-1}}$, while treating the chemical potential $\mu$, the interlayer bias $V_i$ and the damping rate $\eta$ as fitting parameters.
As shown in Fig.~\ref{fig:SNOM_parameters}, changes to these three parameters have drastic effects on the resulting s-SNOM signal around the domain wall.
An increase in $\mu$ increases the plasmon wavelength and decreases the strength of the signal,
a change to $V_i$ changes the signal strength at the wall,
while an increase in $\eta$ decreases the overall amplitude of the oscillations.
This shows that one can reliably determine these three parameters in the fitting procedure.

\begin{figure}[ht]
	\begin{center}
		\includegraphics[width=1.55 in]{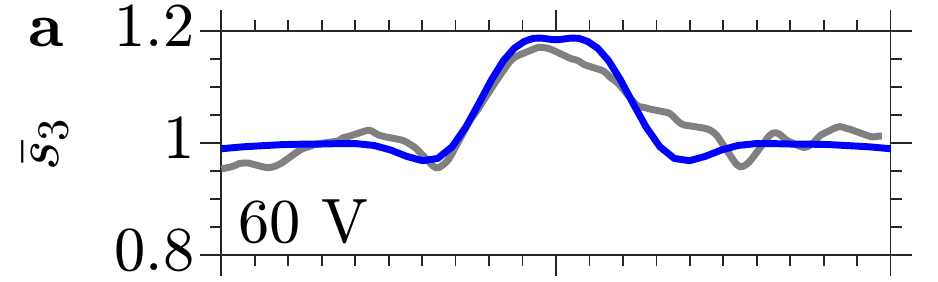}
		\includegraphics[width=1.55 in]{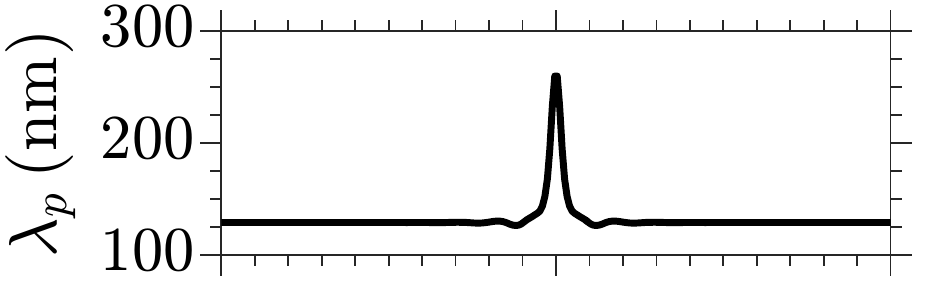}
		\includegraphics[width=1.55 in]{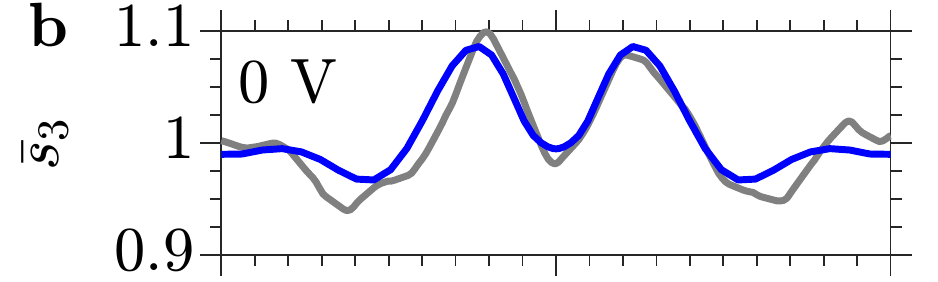}
		\includegraphics[width=1.55 in]{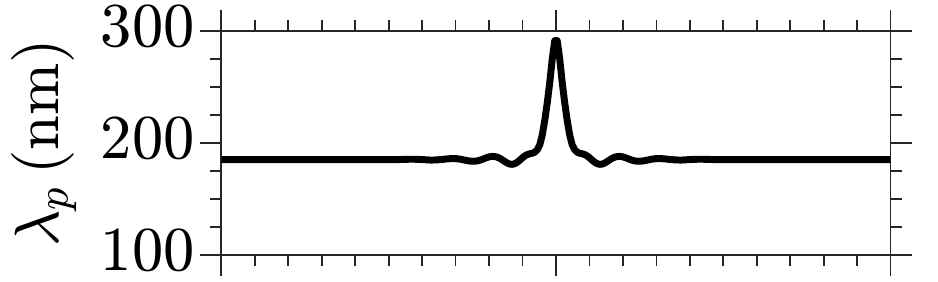}
		\\
		\includegraphics[width=1.55 in]{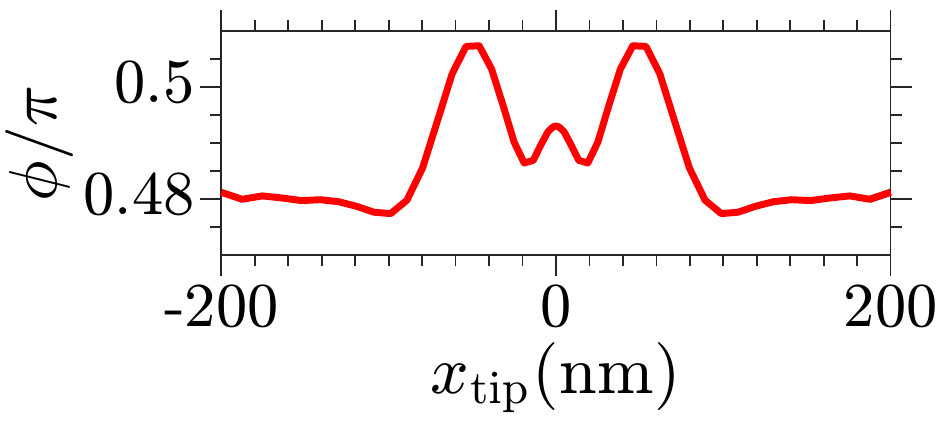}
		\includegraphics[width=1.55 in]{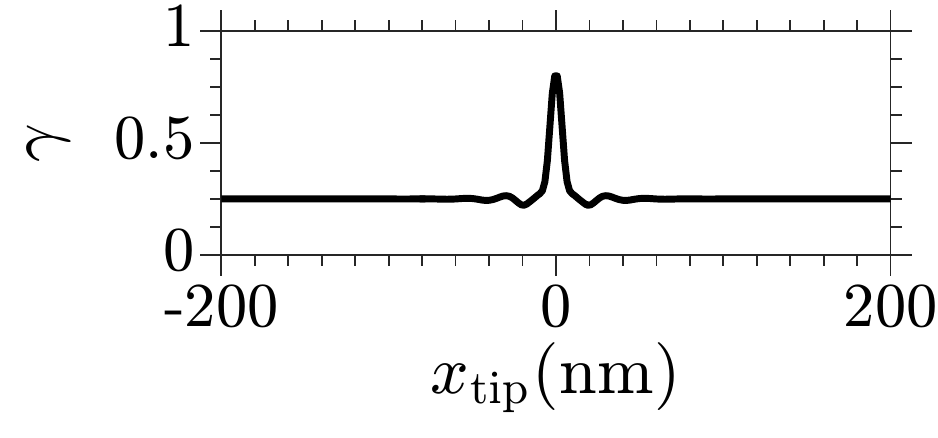}
		\includegraphics[width=1.55 in]{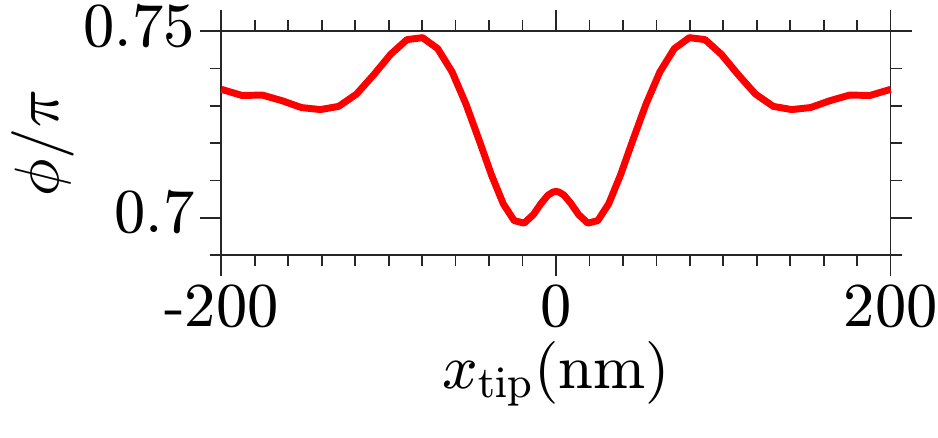}
		\includegraphics[width=1.55 in]{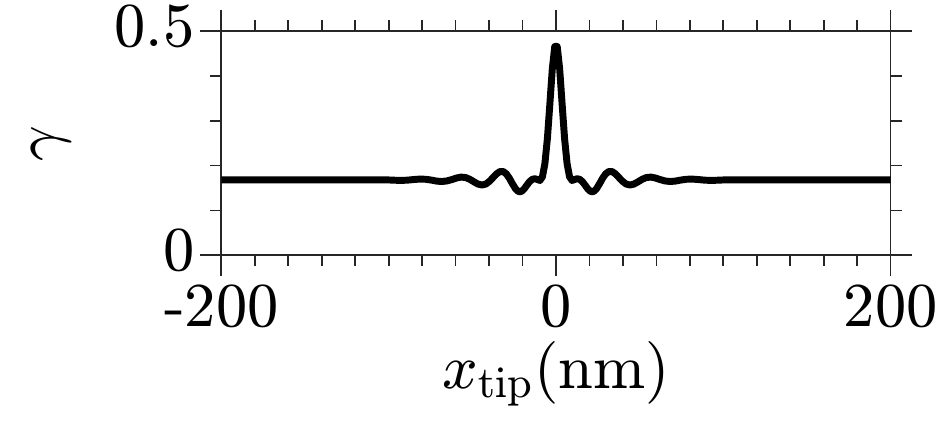}
		\\
		\includegraphics[width=1.55 in]{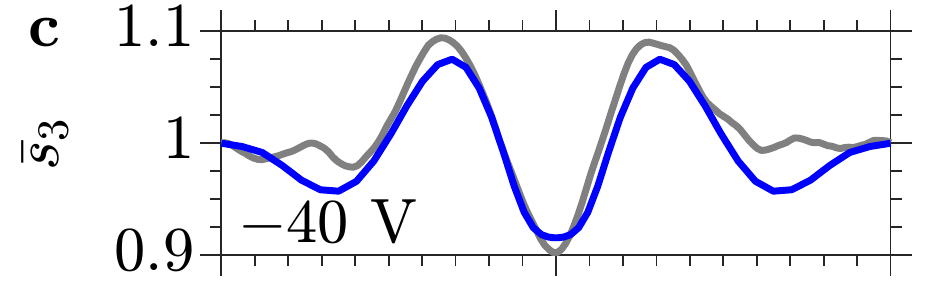}
		\includegraphics[width=1.55 in]{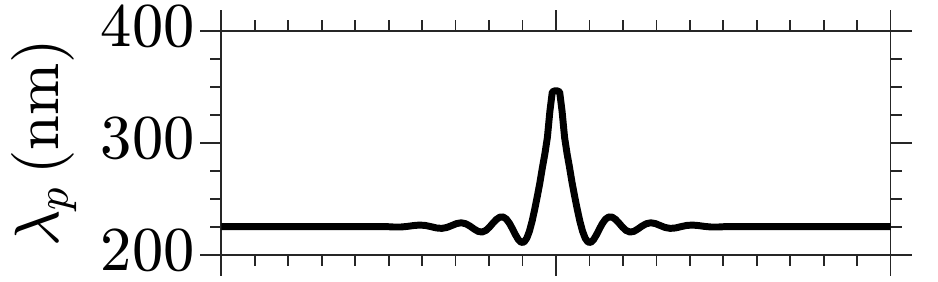}
		\includegraphics[width=1.55 in]{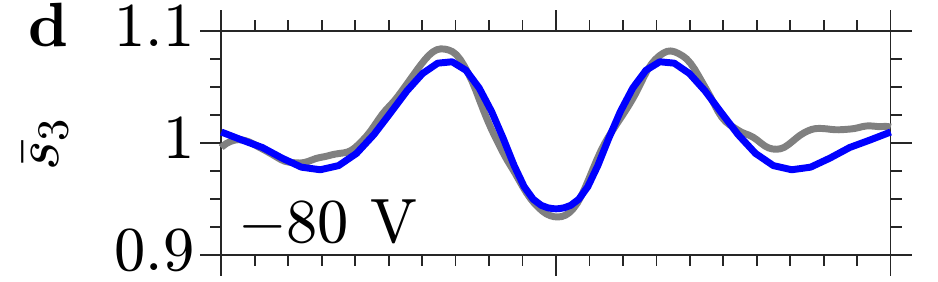}
		\includegraphics[width=1.55 in]{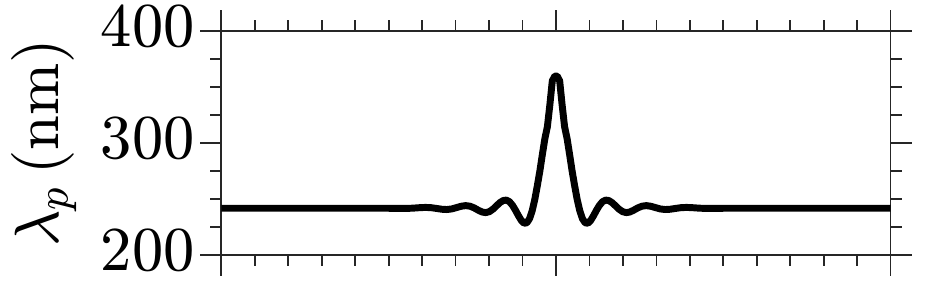}		
		\\
		\includegraphics[width=1.55 in]{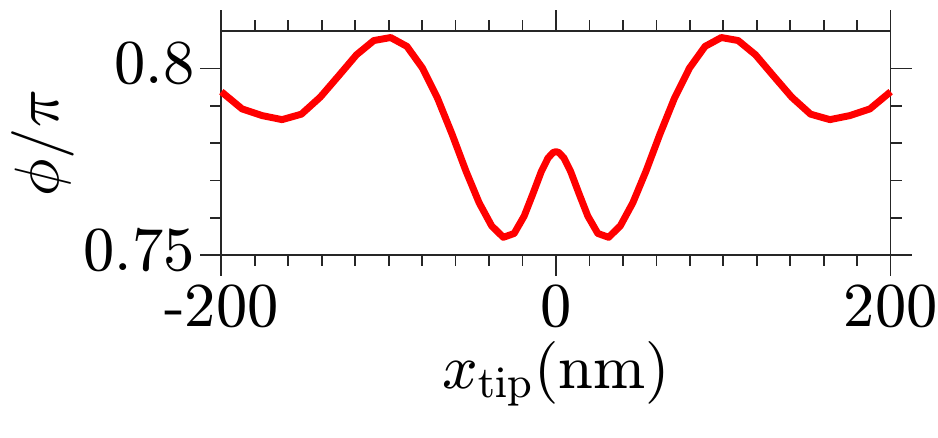}
		\includegraphics[width=1.55 in]{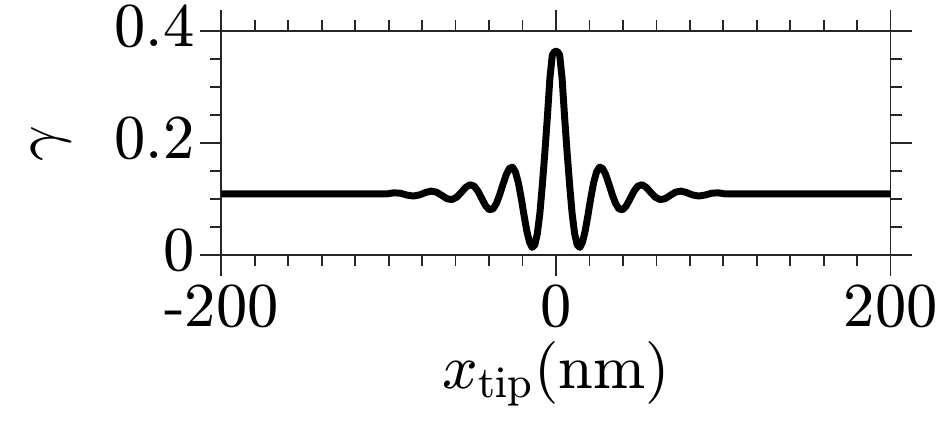}		
		\includegraphics[width=1.55 in]{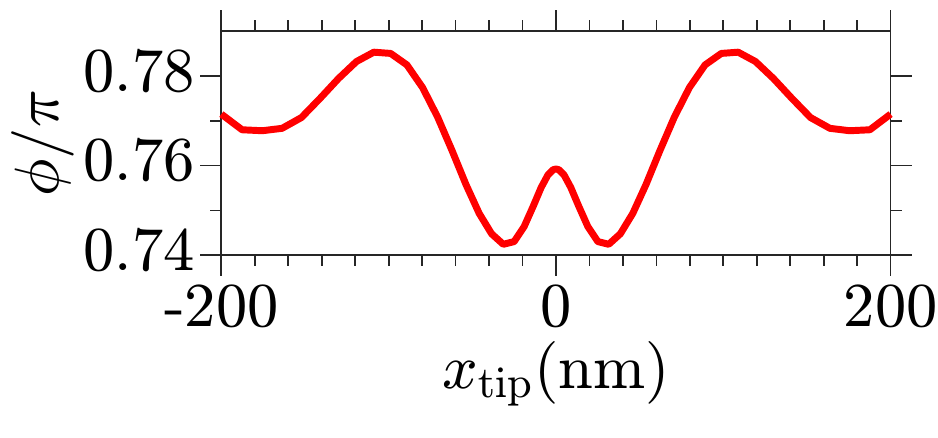}
		\includegraphics[width=1.55 in]{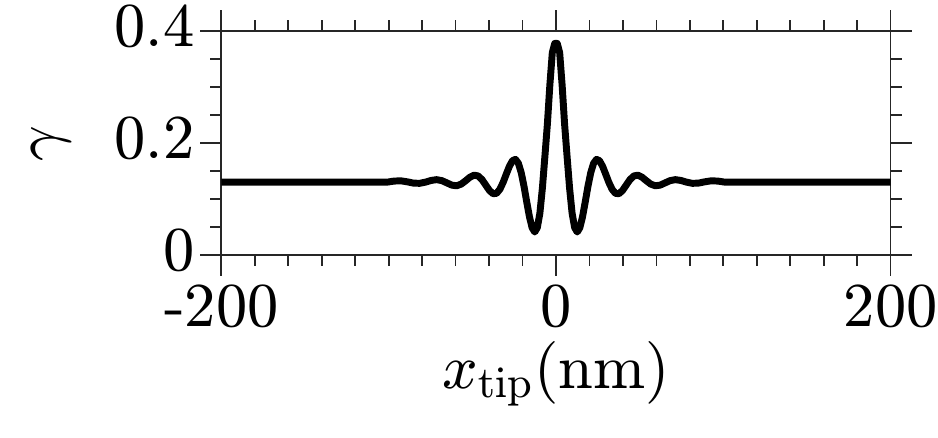}		
	\end{center}
	\caption{Fits for the near-field profiles for the tensile wall. \textbf{a}. $V_g=60\,\mathrm{V}$.
	\textbf{b}. $V_g=0\,\mathrm{V}$.
	\textbf{c}. $V_g=-40\,\mathrm{V}$.
	\textbf{d}. $V_g=-80\,\mathrm{V}$.
	In each panel the normalized experimental near-field amplitude profile $\bar{s}_3$ is shown in gray, the simulated amplitude $\bar{s}_3$ and phase $\phi$ profiles are shown in blue and red.
	Also shown are the plasmon wavelength profile $\lambda_p$ and damping profile $\gamma$ used for the fit.
	}
	\label{fig:SNOM_tensile}
\end{figure}

\begin{figure}[ht]
	\begin{center}
		\includegraphics[width=1.55 in]{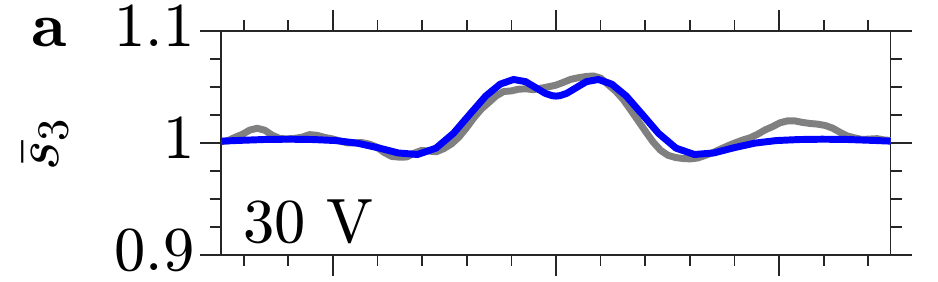}
		\includegraphics[width=1.55 in]{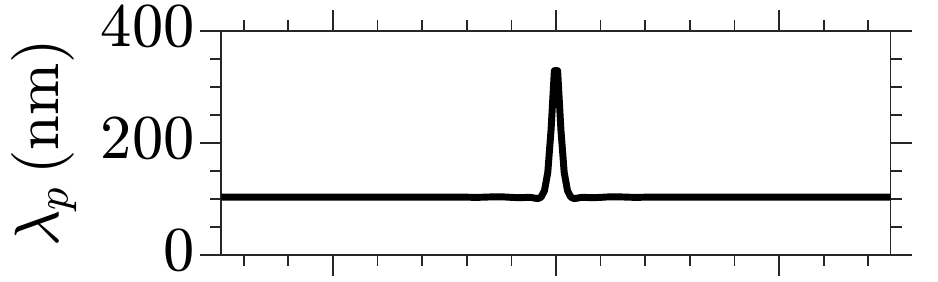}
		\includegraphics[width=1.55 in]{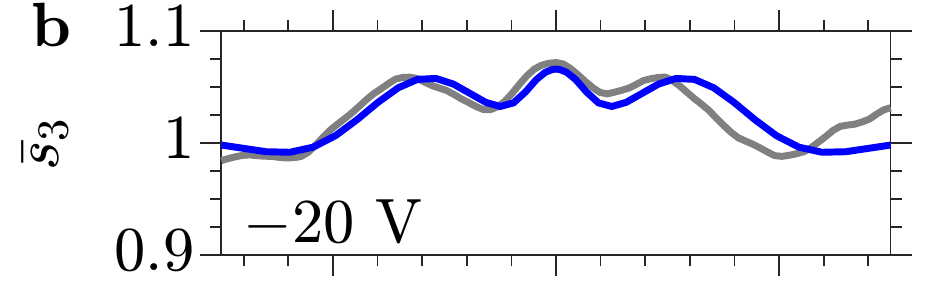}
		\includegraphics[width=1.55 in]{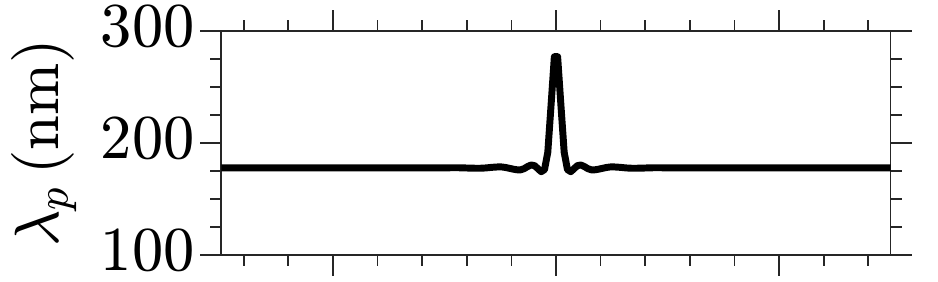}
		\\
		\includegraphics[width=1.55 in]{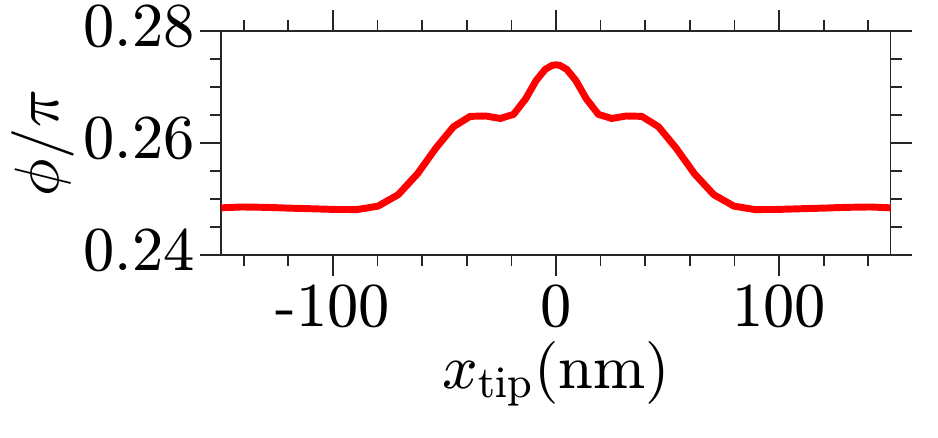}
		\includegraphics[width=1.55 in]{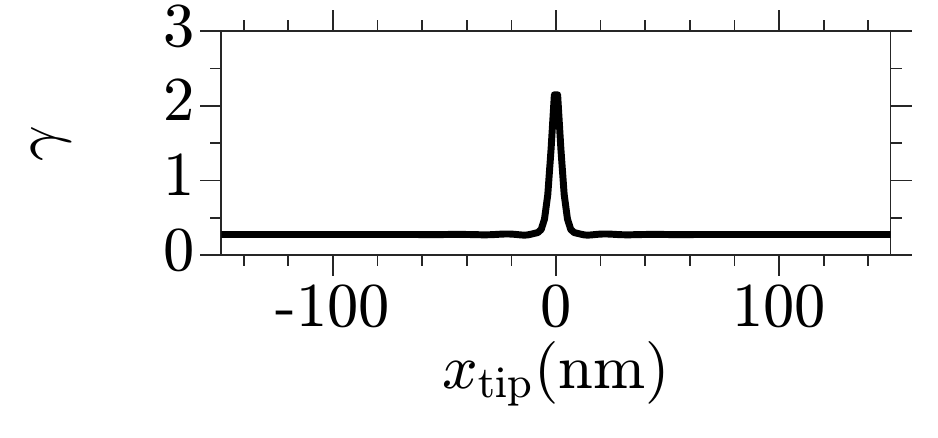}		
		\includegraphics[width=1.55 in]{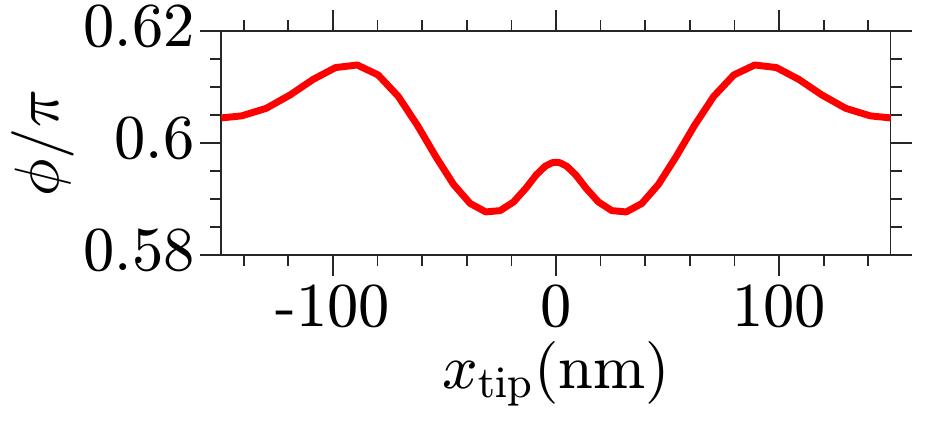}
		\includegraphics[width=1.55 in]{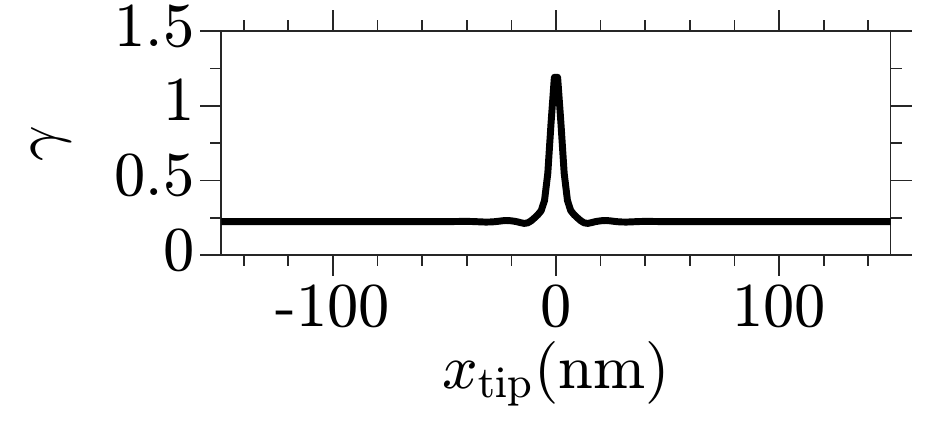}
		\\
		\includegraphics[width=1.55 in]{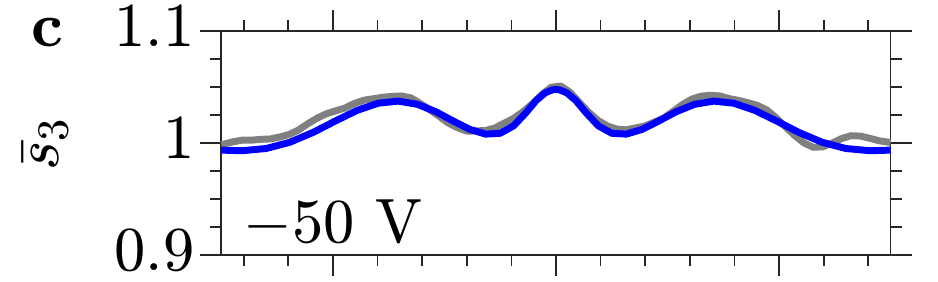}
		\includegraphics[width=1.55 in]{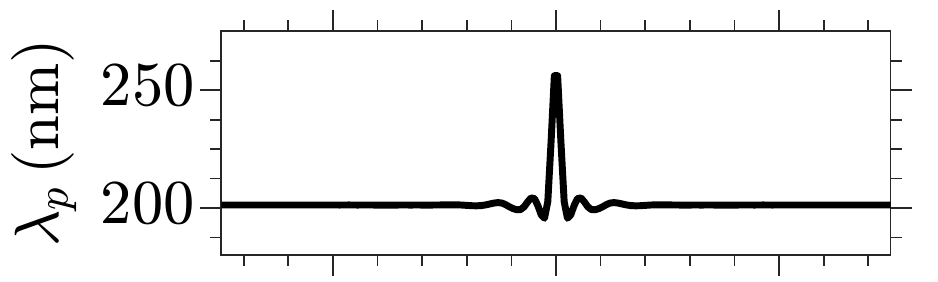}	
		\includegraphics[width=1.55 in]{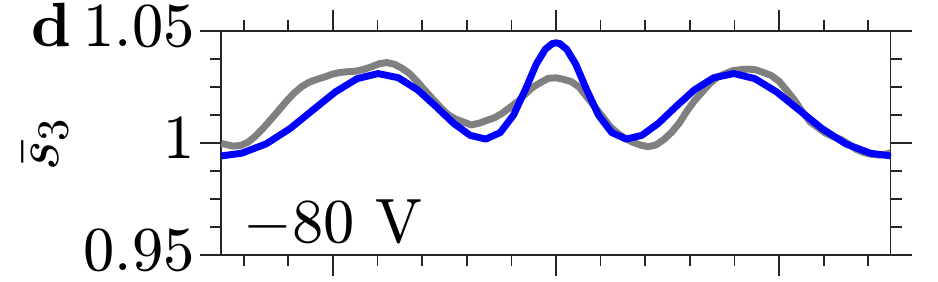}
		\includegraphics[width=1.55 in]{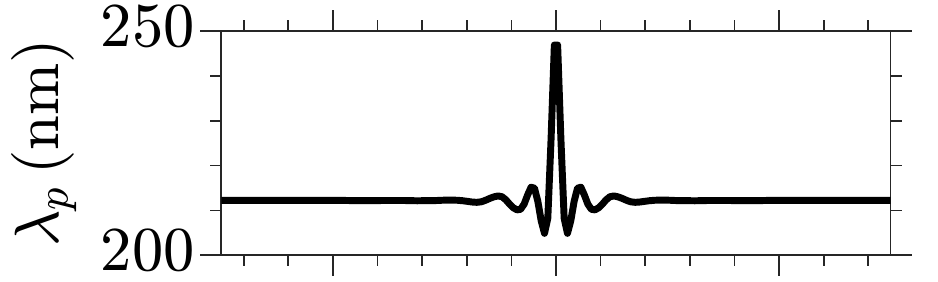}		
		\\
		\includegraphics[width=1.55 in]{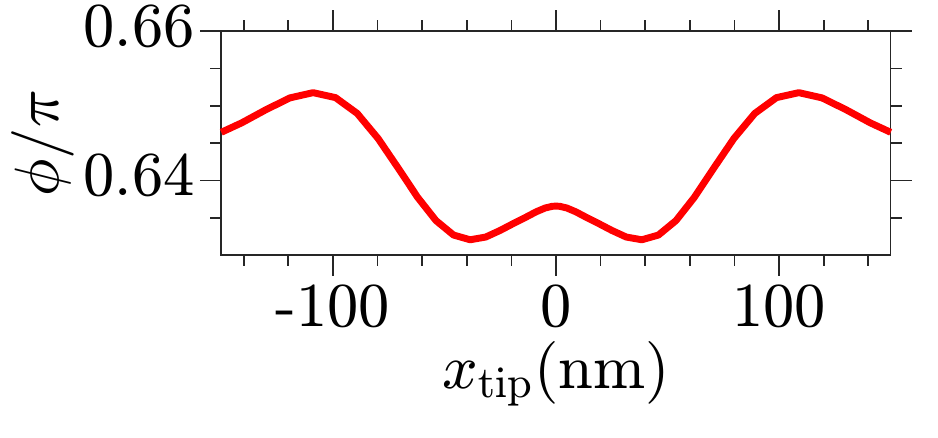}
		\includegraphics[width=1.55 in]{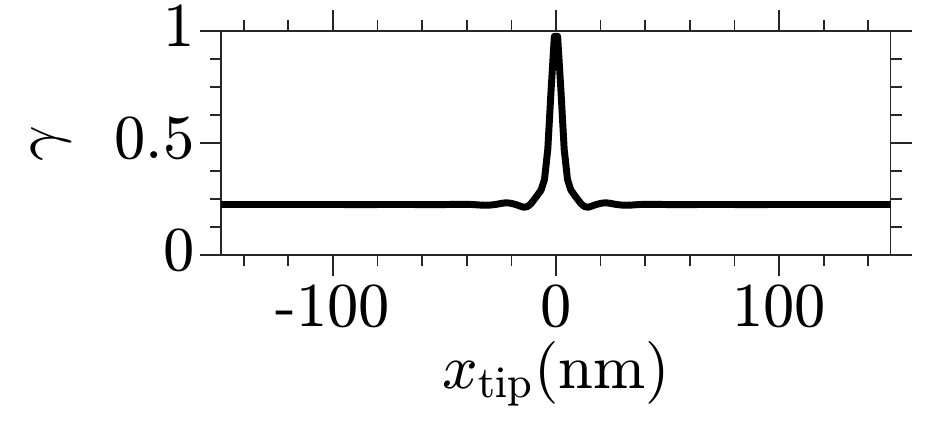}				
		\includegraphics[width=1.55 in]{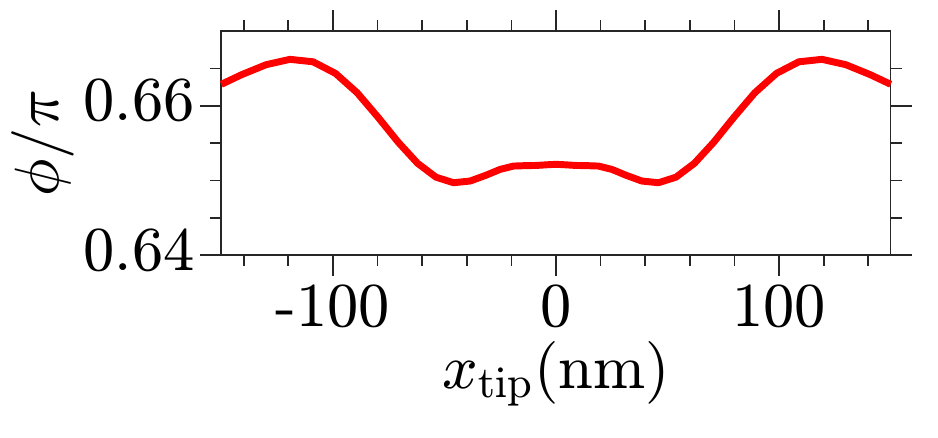}
		\includegraphics[width=1.55 in]{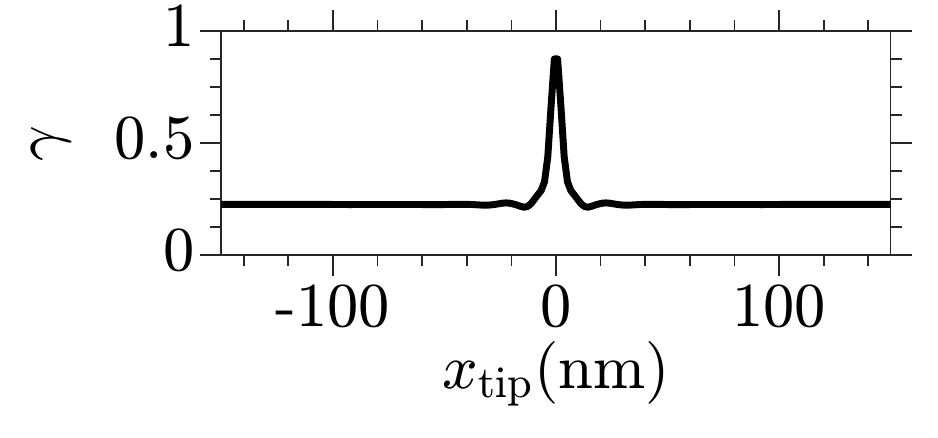}	
		\\	
		\includegraphics[width=1.55 in]{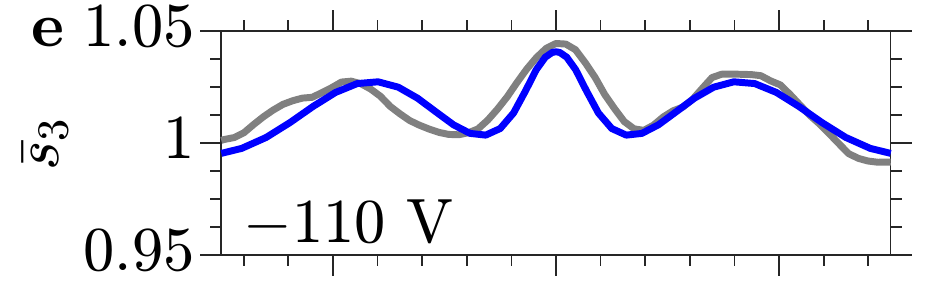}
		\includegraphics[width=1.55 in]{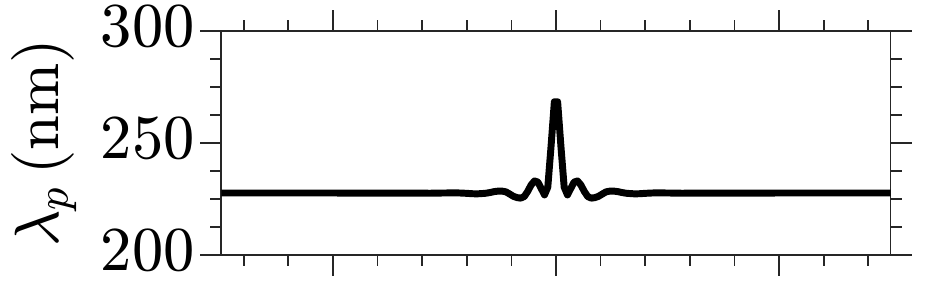}			
		\\
		\includegraphics[width=1.55 in]{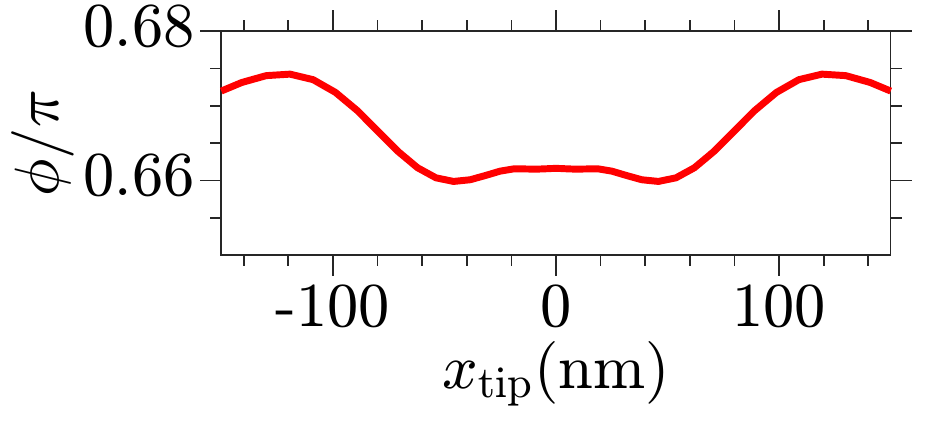}
		\includegraphics[width=1.55 in]{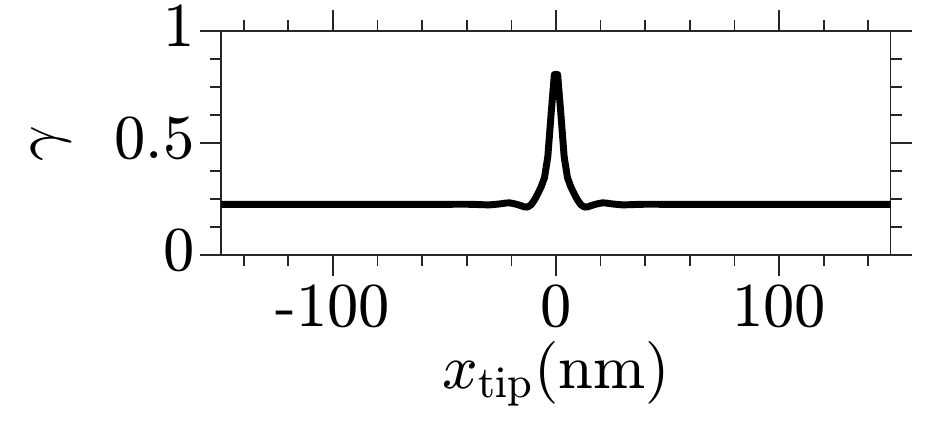}	
	\end{center}
	\caption{Fits for the near-field profiles for the shear wall. \textbf{a}. $V_g=30\,\mathrm{V}$.
	\textbf{b}. $V_g=-20\,\mathrm{V}$.
	\textbf{c}. $V_g=-50\,\mathrm{V}$.
	\textbf{d}. $V_g=-80\,\mathrm{V}$.
	\textbf{e}. $V_g=-110\,\mathrm{V}$.
	In each panel the normalized experimental near-field amplitude profile $\bar{s}_3$ is shown in gray, the simulated amplitude $\bar{s}_3$ and phase $\phi$ profiles are shown in blue and red.
	Also shown are the plasmon wavelength profile $\lambda_p$ and damping profile $\gamma$ used for the fit.
	}
	\label{fig:SNOM_shear}
\end{figure}

In Fig.~\ref{fig:SNOM_tensile} we show our fits to the experimental near-field amplitude profiles for the tensile wall along with the phase $\phi$ of the s-SNOM signal.
Also shown are the plasmonic wavelength profile $\lambda_p$ and the plasmonic damping profile $\gamma$ used for the fit.
Parameters used for the series of fits for $V_g=(60,0,-40,-80)\,\mathrm{V}$ are:
$\mu=(0.17,0.21,0.25,0.27)\,\mathrm{eV}$,
$V_i=(0.25,0.2,-0.1,-0.2)\,\mathrm{V}$,
and $\eta=(0.2,0.15,0.1,0.12)$.
Fits for the shear wall are shown in Fig.~\ref{fig:SNOM_shear}.
The parameters used for $V_g=(30,-20,-50,-80,-110)\,\mathrm{V}$
are:
$\mu=(0.16,0.21,0.23,0.24,0.25)\,\mathrm{eV}$,
$V_i=(0.3,0.35,0.35,0.35,0.4)\,\mathrm{V}$,
and $\eta=(0.2,0.2,0.2,0.2)$.

\section{Dielectric function in the band gap}

In this section we derive the effective 1D dielectric function $\varepsilon_\mathrm{1D}(k_\parallel,\omega)$ of the domain wall
when the chemical potential lies within the band gap. 
The pole of $1/\varepsilon_\mathrm{1D}$ determines the dispersion of the 1D plasmon propagating along the wall.

In the absence of external fields, the total electric potential $\Phi$ of the sheet in the quasistatic limit is determined by the charge density $\rho$ and current density $j$ on the sheet,
\begin{equation}
\Phi=V_2 \ast \rho = V_2 \ast \frac{i}{\omega}\nabla \cdot \mathbf{j}\,,
\label{eqn:phi_j}
\end{equation}
where the Coulomb kernel $V_2=1/\kappa r$, $\mathbf{r}=(x,y)$ and $\ast$ denotes convolution, $A\ast B=\int A(r-r')B(r')dr'$.
For ease of notation we assume that the domain wall lies on the $y$-axis.
When the chemical potential is in the gap, $|\mu|<V/2$ (and the temperature and frequency are low, $k_B T\ll V$ and $\hbar \omega\ll V$), 
only the bound states contribute to the optical response.
The charge density is zero on the sheet and the current only flows along the domain wall,
so we can make the simplification $j_x=0$ and $\Phi=\phi(x)e^{iq_y y}$.
Eq.~\eqref{eqn:phi_j} can then be rewritten as
\begin{equation}
\phi(x) = V_{1} \ast \frac{i}{\omega}\partial_y j_y = V_{1} \ast \frac{k_y^2}{i\omega}\int\Sigma_{yy}(x,x')\phi(x')dx'\,,
\label{eqn:phi_1D}
\end{equation}
Here the 1D Coulomb kernel is $V_1(x)=\int V_2 dy=\frac2\kappa K_0(q_y|x|)$, where $K_0$ is the modified Bessel function of the second kind.
Note that we removed the $k_y$ dependence in $\Sigma_{yy}$ as the plasmon wavelength $\lambda_y\sim k_y^{-1}$ is much larger than all other length scales in the problem, so that we can make the approximation $\Sigma(x,x',k_y)\simeq \Sigma(x,x',0)$.

At frequencies $\hbar \omega\ll V$, there are no allowed optical transitions and the conductivity comes purely from the Drude response,
\begin{equation}
\Sigma_{yy}(x,x')=g_s\sum_{K,K'}\sum_{j=1}^{N} \frac{iD_{yy,j}}{\pi(\omega-\xi_j v_j q_y)}|\psi_j(x)|^2|\psi_j(x')|^2\,,
\end{equation}
where $\psi_j$ is the wavefunction of the $j$-th bound state at energy $E_j=\mu$ and $g_s=2$ is the spin degeneracy.
The Drude weight $\frac1\pi D_{yy,j}=\frac{e^2}{h}|v_j|$ is directly proportional to the particle group velocity $v_j=\partial E_j/\hbar\partial k_y$, and $\xi_j$ is the sign of $v_j$.
The summation over the $K$ and the $K'$ valleys can be reduced by noting that
every bound state has a counterpart in the other valley with a velocity that is equal in magnitude but opposite in direction.

Since the width of the wavefunctions $\sim l$ is much smaller than the plasmon wavelength $\lambda_y$, the particle density distributions $|\psi_j(x)|^2$ can be roughly approximated as a $\delta$-function of characteristic width $l$.
Eq.~\eqref{eqn:phi_1D} then becomes
\begin{equation}
\phi(l)\simeq\left(\sum_{j=1}^{N}\frac{k_y^2}{\kappa(\omega^2-v_j^2k_y^2)}2K_0(k_yl)g_s\frac{2e^2}{h}|v_j|\right)\phi(l)=\left(1-\frac{\varepsilon_{1D}}{\kappa}\right)\phi(l)\,.
\end{equation}
For small arguments $K_0(z)\simeq \log(A/z)$ where $A\simeq 2e^{-0.577}=1.12$, and so the 1D dielectric function is
\begin{equation}
\varepsilon_{\mathrm{1D}}\left(k_y, \omega\right) = 
\kappa - \frac{8 e^2}{h}\, \ln \left(\frac{A}{k_y l}\right)
k_y^2\sum_{j = 1}^{N}
\frac{ |v_{j}|}{\omega^2 - k_y^2 v_{j}^2}\,.
\label{eqn:1D_plasmon}
\end{equation}


\bibliographystyle{naturemag}
\bibliography{abba_plasmon}